\documentclass[a4paper,fleqn]{cas-dc}

\usepackage{amsmath}
\usepackage{wasysym} 

\usepackage{physics}
\usepackage{url}
\usepackage[T1]{fontenc} 
\usepackage{textcomp,textgreek}
\usepackage[authoryear,longnamesfirst]{natbib}

\def\tsc#1{\csdef{#1}{\textsc{\lowercase{#1}}\xspace}}
\tsc{WGM}
\tsc{QE}

\usepackage{derivative}
\usepackage{graphicx}
\usepackage{tikz}
\usetikzlibrary{shapes,arrows}

\usepackage{enumitem}
\newlist{steps}{enumerate}{1}
\setlist[steps, 1]{label = Step \arabic*:}
\usepackage{float, gensymb}
\usepackage{booktabs}

\newcommand{\newandsetlength}[2]{\newlength{#1}\setlength{#1}{#2}}
\newandsetlength{\fwidth}{7.5cm}
\newandsetlength{\fdouble}{7.5cm}

\usepackage{framed} 
\usepackage{multicol} 
\usepackage{nomencl} 
\makenomenclature
\renewcommand*\nompreamble{\begin{multicols}{2}}
\renewcommand*\nompostamble{\end{multicols}}
\usepackage{etoolbox}
\renewcommand\nomgroup[1]{%
  \item[\bfseries
  \ifstrequal{#1}{N}{Notations}{%
  \ifstrequal{#1}{S}{Subscripts and superscripts}{%
  \ifstrequal{#1}{A}{Abbreviations}{}}}%
]}

\begin{document}
\let\WriteBookmarks\relax
\def\floatpagepagefraction{1}
\def\textpagefraction{.001}

\shorttitle{Modeling and Control of Heat Exchangers}
\shortauthors{Skantzikas et al.}

\title[mode=title]{Physics-informed neural control of heat flow rate in a counter-flow heat exchanger: modeling, identification and experimental validation}

\author[1]{Konstantinos Skantzikas}[orcid=0009-0009-7344-4722]
\ead{kostas.skantzikas@gipsa-lab.fr}
\credit{Methodology, Software, Investigation, Writing -- original draft}

\author[1]{Emmanuel Witrant}[orcid=0000-0002-2026-8915]
\cormark[1]
\ead{emmanuel.witrant@univ-grenoble-alpes.fr}
\credit{Conceptualization, Supervision, Funding acquisition, Writing -- review and editing}

\author[2]{Bojan Mavkov}[orcid=0000-0003-4812-8049]
\ead{mavkov@i3s.unice.fr}
\credit{Methodology, Supervision, Funding acquisition, Writing -- review and editing, Validation}

\affiliation[1]{
organization={Université Grenoble Alpes, CNRS, GIPSA-lab},
addressline={11 Rue des Mathématiques},
city={Saint-Martin-d'Hères},
postcode={38402},
country={France}
}

\affiliation[2]{
organization={Université Côte d'Azur, CNRS, I3S},
addressline={CS 40121},
city={Sophia Antipolis Cedex},
postcode={06903},
country={France}
}

\cortext[1]{Corresponding author.}

\begin{abstract}
The advanced operation of heat exchangers plays a critical role in the development of efficient and sustainable multivector energy systems. This paper proposes a hybrid physics-informed and data-driven framework for the analysis, identification, and regulation of transient heat transfer in counter-flow heat exchangers. A parameterized physical model combines classical heat-transfer relations with transport equations to capture convective propagation, transport-induced delays, and nonlinear effects associated with flow-rate variations. The heat-transfer coefficient is identified from experimental data, enabling the characterization of flow-dependent heat-transfer dynamics over a broad operating range.

The identified model is then used to develop a hierarchical real-time control strategy. An inner loop regulates the volumetric flow rate using a high-order data-driven linear model, while an outer loop regulates the exchanged heat power using a machine-learning controller initially trained on the physical model and subsequently refined with experimental data. 
Experimental results demonstrate accurate reproduction of transient heat-transfer dynamics and effective heat-flow-rate regulation under varying thermal boundary conditions (identification fit above 94\% in normalized root-mean-square error and total variation of the flow rate reduced by half compared to a proportional-integral controller). The proposed framework combines physical interpretability with data-driven adaptation for the analysis and real-time operation of dynamic heat-transfer systems.
\end{abstract}

\begin{graphicalabstract}
\includegraphics[width=1.\linewidth]{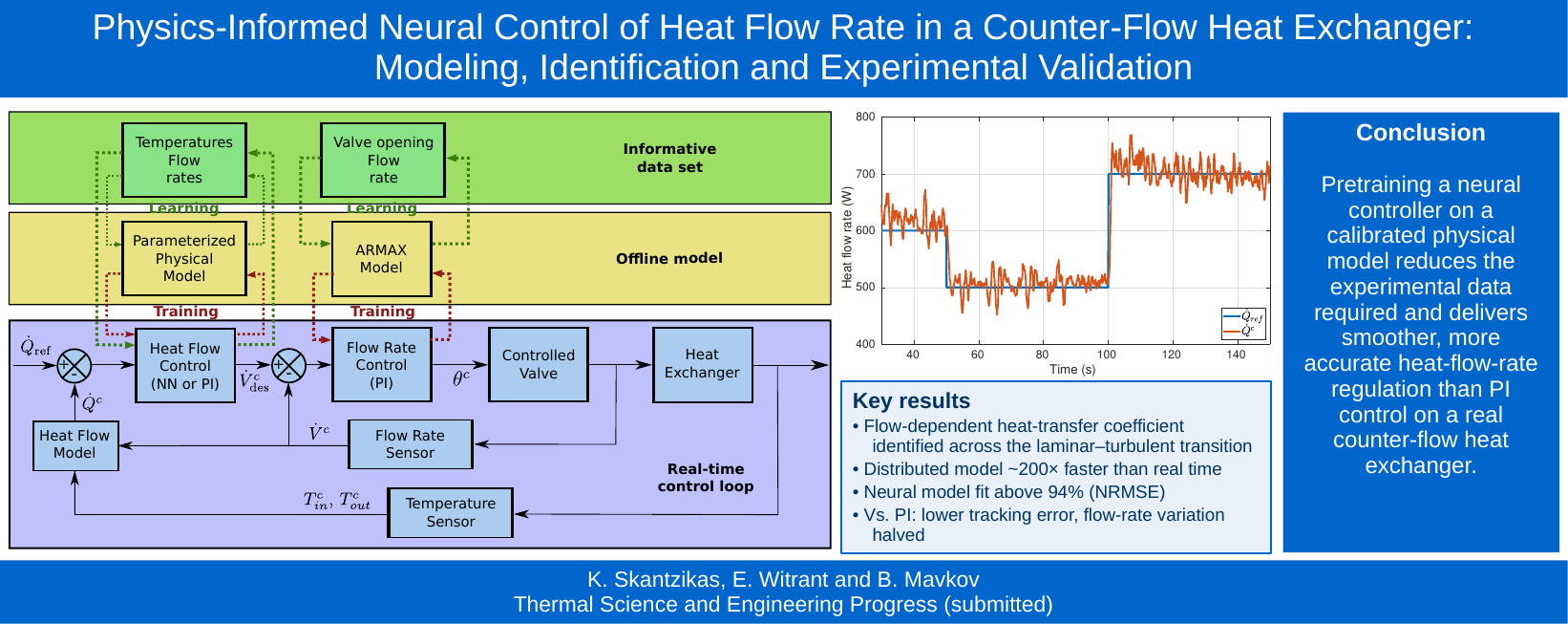}
\end{graphicalabstract}

\begin{highlights}
\item Distributed physics-based model captures convective delays and nonlinear heat flow
\item Flow-dependent heat-transfer coefficient identified across laminar--turbulent regimes
\item Physics-based pretraining reduces experimental data needed for neural control design
\item Neural controller improves tracking and smoothness over proportional--integral control
\item Framework combines physical interpretability with real-time data-driven adaptation
\end{highlights}


\maketitle




\begin{table*}[!t]   

\begin{framed}

\nomenclature[N]{$A_c$}{Free flow area (m$^2$)}
\nomenclature[N]{$C_p$}{Specific heat (J/kg.K)}
\nomenclature[N]{$d_H$}{Hydraulic (inner) diameter of the inner tube (m)}
\nomenclature[N]{$h$}{Heat transfer coefficient (W/m$^2$.K)}
\nomenclature[N]{$k$}{Thermal conductivity (W/m.K)}
\nomenclature[N]{$L$}{Length of HE tube (m)}
\nomenclature[N]{$m$}{Mass (kg)} 
\nomenclature[N]{$p$}{Inner perimeter of the inner tube (m)}
\nomenclature[N]{$\dot{Q}$}{Heat flow rate (W)}
\nomenclature[N]{$S$}{Area (m$^2$)}
\nomenclature[N]{$S^{he}$}{Heat exchange surface (m$^2$)}
\nomenclature[N]{$t$}{Time (s)}
\nomenclature[N]{$t_s$}{Sampling time (s)}
\nomenclature[N]{$T$}{Temperature (K)}
\nomenclature[N]{$V$}{Volume (m$^3$)}
\nomenclature[N]{$\dot{V}$}{Volumetric flow rate (m$^3$/s)}
\nomenclature[N]{$x$}{Position along the HE (m)}
\nomenclature[N]{$z$}{$\mathcal{Z}$-transform operator}
\nomenclature[N]{$\Delta{T}$}{Average temperature difference (K)}
\nomenclature[N]{$\mu$}{Water viscosity (Pa.s)} 
\nomenclature[N]{$\rho$}{Density (kg/m$^3$)}
\nomenclature[N]{$\sigma$}{Standard deviation of the signal}
\nomenclature[N]{$\theta$}{Position signal sent to the valve}

\nomenclature[S]{$a$}{Ambient air}
\nomenclature[S]{$c$}{Cold fluid}
\nomenclature[S]{$\text{des}$}{Desired value}
\nomenclature[S]{$h$}{Hot fluid}
\nomenclature[S]{$he$}{Heat exchange}
\nomenclature[S]{$in$}{Inlet}
\nomenclature[S]{$inox$}{316L stainless steel}
\nomenclature[S]{$LM$}{Log-Mean Temperature Difference}
\nomenclature[S]{$out$}{Outlet}
\nomenclature[S]{$\text{ref}$}{Reference value}
\nomenclature[S]{$SS$}{Steady-state value}
\nomenclature[S]{$tank$}{Cold tank}
\nomenclature[S]{$w$}{Water}
		
\nomenclature[A]{ARMAX}{Auto-Regressive with Moving Average and eXogenous input}
\nomenclature[A]{ETFE}{Empirical Transfer Function Estimate}
\nomenclature[A]{HE}{Heat exchanger}
\nomenclature[A]{LMTD}{Log-Mean Temperature Difference}
\nomenclature[A]{MSE}{Mean square error}
\nomenclature[A]{NARMA-L2}{Nonlinear AutoRegressive Moving Average}
\nomenclature[A]{Nu}{Nusselt number}
\nomenclature[A]{PDE}{Partial differential equation}
\nomenclature[A]{PI}{Proportional-Integral}
\nomenclature[A]{Pr}{Prandtl number}
\nomenclature[A]{PRBS}{Pseudo-Random Binary Sequence}
\nomenclature[A]{Re}{Reynolds number}
\nomenclature[A]{STDV}{Standard deviation}
\nomenclature[A]{TV}{Total variation}

\printnomenclature

\end{framed}

\end{table*}

\section{Introduction}

Industrial energy losses often reach 20 to 50\% of the energy consumed \citep{BCS2008}.
Heat exchangers (HE) are key components of multivector energy systems, where their operating conditions may vary significantly in response to fluctuating thermal loads and energy demands.
While conventional heat-transfer analysis primarily relies on steady-state formulations, dynamic operation requires an accurate description of transient transport phenomena and their impact on heat-transfer performance.
In this context, the ability to predict and regulate the exchanged heat power under varying inlet conditions is essential for efficient thermal management and heat recovery.
By introducing artificial intelligence to analyze and adapt the operation of these systems in real time, companies can increase their energy efficiency and minimize their ecological impact.
Extensive research has been conducted on the design and optimization of HEs to improve energy efficiency under steady-state conditions. 
However, transient phenomena, such as pressure drops, variations in thermal load, or phase changes of the working fluids, are rarely studied due to the complexity of their nonlinear and distributed dynamics. 
In practical applications involving fluctuating thermal loads, intermittent renewable energy sources, or dynamic process operation, transient transport phenomena may become dominant and strongly affect thermal performance. 
Understanding and predicting these effects therefore remains an important challenge in heat-transfer research, particularly because the interaction between convective transport, spatial temperature gradients, and time-varying boundary conditions is not fully captured by classical lumped models.
A key research direction in HE optimization is the operation and control of dynamic behavior, particularly in the presence of disturbances and transient conditions, e.g. to ensure that a reference heat flow rate is transferred by the exchanger despite varying operating conditions.  
Controllability, dynamic performance, and diagnostic criteria are thus at the forefront of the operability challenges of HEs~\citep{CHEW201317}, whether for upgrading existing installations or designing new ones.

Dynamic modeling of HEs has been a longstanding focus and motivated several control approaches. Early works based on \textit{first-principles modeling} date back to the 1950s, including the transient analysis of lumped dynamics and state prediction~\citep{Takahashi1952, Mozley1956}.  
Lumped linear models were later simplified through the linear cross-convection model for predictive control by~\cite{laszczyk1999application}, and further extended to include mass flow rate effects under steady flow conditions via parameterization by~\cite{LASZCZYK2017140}.  
Several control approaches based on lumped linear dynamics have been successfully implemented in HE experiments.  
For instance, proportional-integral-derivative (PID) control has been combined with feedforward control~\citep{Khare_2010}, fuzzy logic controllers~\citep{ALDAWERY20126873}, and relay auto-tuning and internal model control methods using an auto-regressive with moving average and exogenous input (ARMAX) representation~\citep{SAHOO2017124}.  
The existence and uniqueness of solutions for lumped dynamics of compartmental models with constant flow rates are analyzed by~\cite{ZAVALARIO20071739}, along with a stability analysis of the equilibrium states.
This stability property is exploited by~\cite{ZAVALARIO2009136} to design a controller for flow regulation that only needs a partial state knowledge.
Model predictive control techniques for lumped models have also been explored, both for individual units~\citep{DYRSKA2023120366} and for networks of HEs~\citep{JALAL2024121441}.

More advanced dynamic modeling and control methods account for complex physical processes that involve nonlinearities, uncertainties, and spatially distributed phenomena.
The temperature dynamics within the HE can be described by a system of coupled, first-order hyperbolic partial differential equations (PDEs). Specifically, applying heat transport equations and conservation of energy, in conjunction with appropriate assumptions for 1-D modeling, leads to the system of first-order PDEs proposed by \cite{ANSARI20062401}, which describes the dynamics of the two fluids and of the wall.
Dividing the wall enthalpy between compartments of the inside and outside fluids, a simplified model is proposed by \cite{bonne:tel-01133432}, where the wall dynamics is included in the hot and cold flows (increasing their effective time constants) and the model is reduced to two states.
This model is used for real-time estimation and observer design by \cite{Hamze_2018,ZOBIRI20177127}, and enables the analysis of thermal wave propagation along the exchanger.
The accuracy of dynamic models is also strongly affected by the heat-transfer coefficient. 
While standard approximations use parameterized expressions depending mostly on the Reynolds number (Re) \citep{Gnielinski75,VERAGARCIA20101231}, considering boundary losses in a more global perspective is also crucial~\citep{KUMAR2026129157}.

Data-based methods and machine learning (ML) algorithms have been increasingly applied to enhance HE efficiency and address maintenance needs (e.g. to prevent fouling, the accumulation of unwanted materials that degrades heat-transfer performance).  
Although many studies focus on offline design optimization, recent research highlights the potential of ML to improve control system adaptability and performance under varying operating conditions.
For example, a neural network (NN) approach for nonlinear system identification and control is presented by~\cite{BITTANTI1997135}, and evaluated on simulation results. \cite{DIAZ20011671} predict the dynamic behavior of an HE using a NN, and a second network controls the temperature of the air passing over it. 
\cite{carvalho2020implementation} use an artificial NN model in a predictive control method to control the outlet oil stream temperature of an HE.
\cite{Ragupathi2024} predict heat transfer in spiral tubes using regularized NNs. 
A physics-informed NN model is used by \cite{majumdar2025hxpinn} for real-time monitoring of an industrial air heater. ML is combined with computational fluid dynamics by~\cite{HE2025113759}, in an online method for rapid fouling detection, leveraging real-time estimation of the heat-transfer coefficient.  
\cite{Najjar2016} utilize a reference quasi-static physical model to train ML algorithms for diagnosis and optimal sensor placement. 
A \emph{Long Short-Term Memory} NN (capable of capturing long-range temporal dependencies) is employed by \cite{HOU2025108809} to predict temperature evolution and system health status, supported by experimental validation.
More related references can be found in the recent survey on NNs for thermal analysis and control of heat exchangers by~\cite{MOHANRAJ2015150}. 
This survey highlights three limitations of NNs that are considered in this paper: overtraining errors (need to optimize the training cycles and number of training data), extrapolation errors (representative information on the validity domain), and training data requirement (a large and optimized data set is necessary).
These limitations are addressed here by first defining a parameterized physical model prior to the NN definition and then selecting a NN architecture that is suitable for efficient feedback control.

In this work, a hybrid physics-informed and data-driven framework is proposed to analyze, identify, and regulate transient heat-transfer mechanisms in counter-flow heat exchangers. Particular attention is devoted to the characterization of transport-induced delays, convective propagation effects, and flow-dependent heat-transfer dynamics that are typically neglected in steady-state formulations.
The main contributions are:
\begin{enumerate}
\item a distributed transient heat-transfer model that combines transport equations with classical heat-exchanger theory and accurately reproduces thermal-wave propagation phenomena;
\item an experimental identification methodology allowing the reconstruction of a flow-dependent heat-transfer coefficient and the derivation of a Nusselt–Reynolds map over a wide operating range;
\item a hybrid machine-learning strategy in which the controller is first trained on the identified physical model and subsequently adapted using experimental data, thereby reducing the amount of experimental data required while accounting for unmodeled dynamics.
\end{enumerate}

The remainder of the paper is organized as follows.  
Section~\ref{sec:experimental} describes the experimental test bench and the proposed control architecture.  
Section~\ref{sec:innerloop} details the inner control loop, which focuses on modeling and controlling the cold fluid's volumetric flow rate using an identified ARMAX model and PI regulation.  
Section~\ref{sec:physical_model} presents the physical model, including heat transport phenomena and parameterized by an estimated flow-dependent heat-transfer coefficient.
The outer control loop, which integrates the parameterized physical model in a machine learning strategy and governs heat-flow-rate regulation, is described in Section~\ref{sec:outerloop}.

\section{Heat Exchanger Test Bench and Control Architecture}

\label{sec:experimental}

\begin{figure*}
\centering
\includegraphics[width=1.\linewidth]{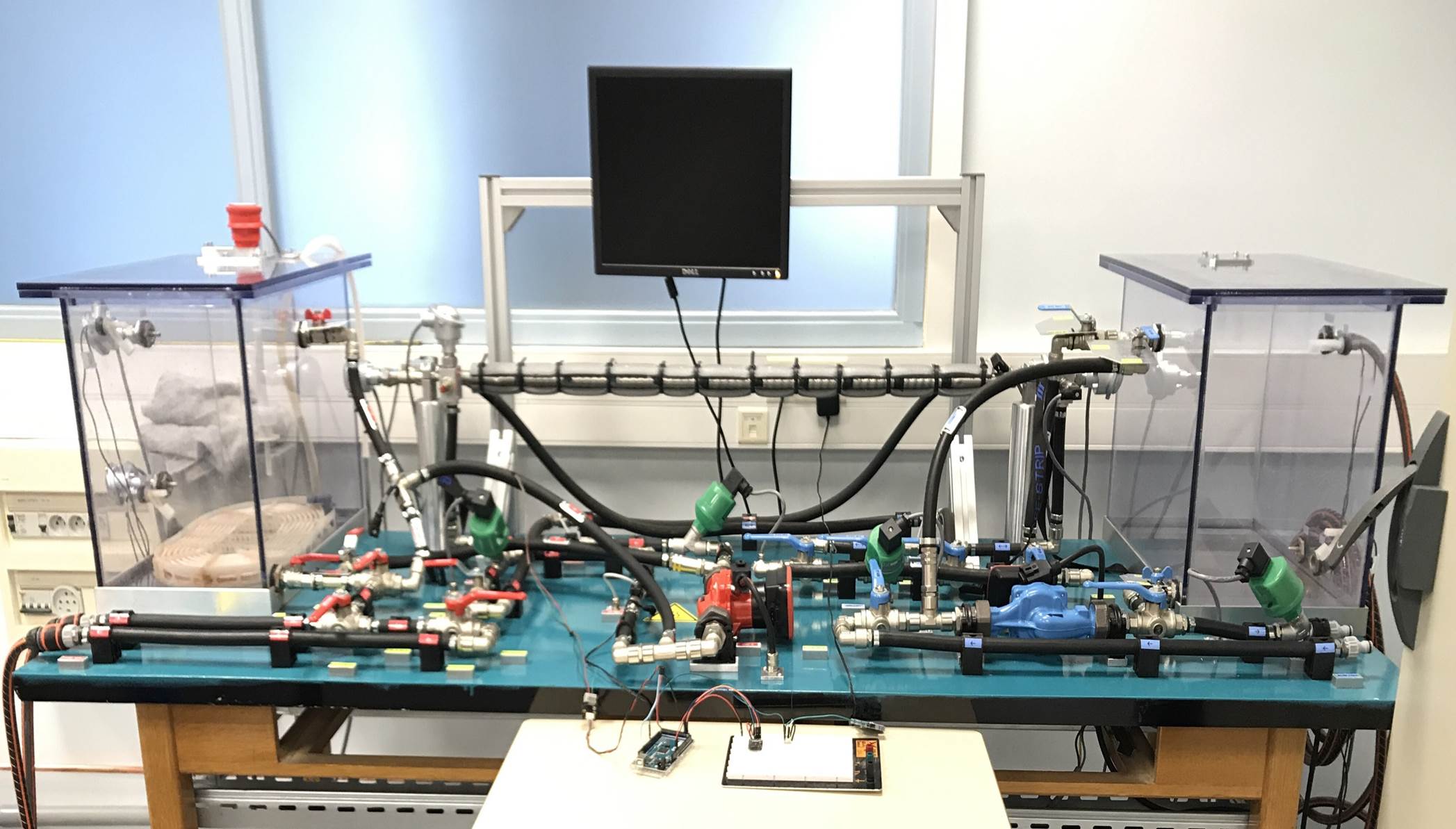}
\caption{Test bench for heat-flow-rate regulation at GIPSA-lab, consisting of two tanks (hot and cold), an HE, water pumps, regulated valves and temperature sensors.}
\label{fig:Gipsa_Heat}
\end{figure*}

An experimental test bench for HE control has been developed at GIPSA-lab to investigate transport phenomena and transient thermal behavior. This test bench is depicted in Fig.~\ref{fig:Gipsa_Heat}, and a technical description of its components is provided in Appendix~\ref{App:TestBench}. This section describes the system configuration and the proposed control architecture.

\subsection{Heat Exchange Setup}

\begin{figure*}
\centering
\includegraphics[width=1\linewidth]{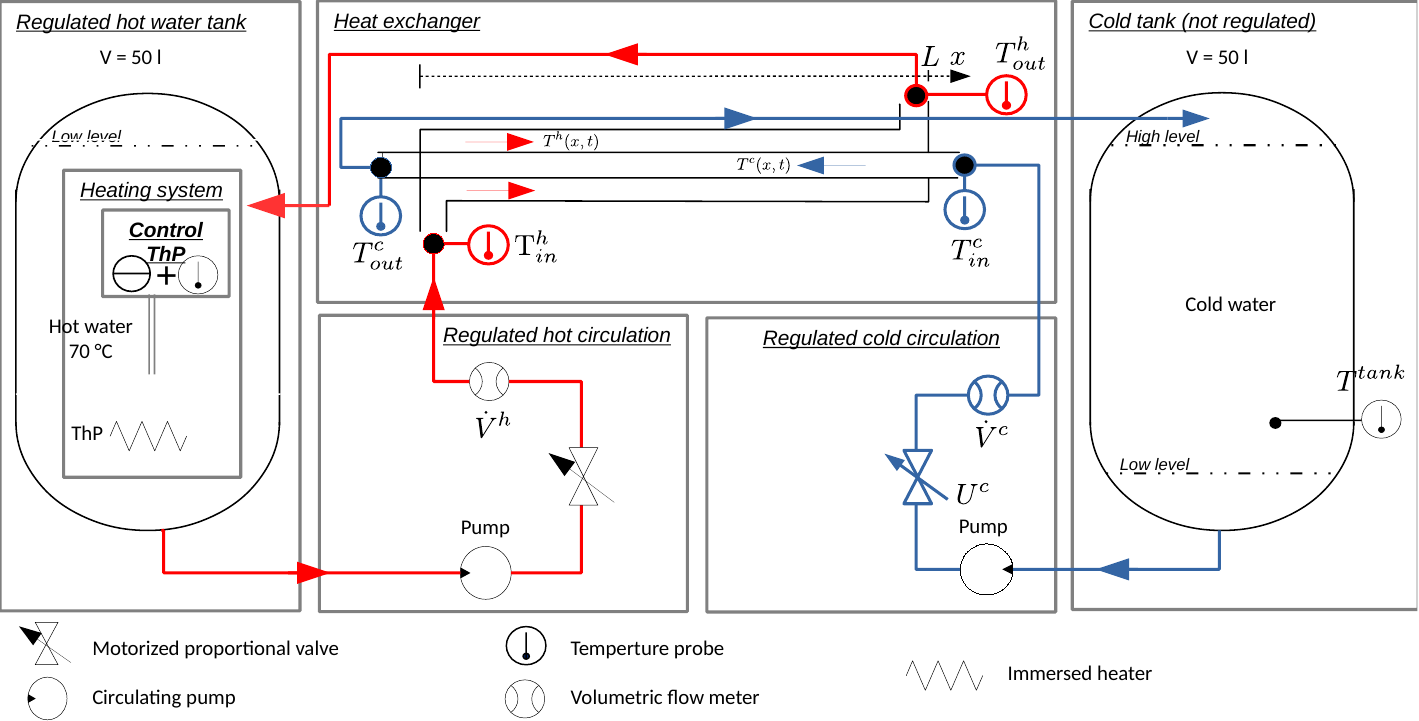}
\caption{Test bench configuration for cold-flow-rate regulation. Sensors and actuators used in this study are indicated by colored lines. Temperatures are measured at the inlets ($T_{in}^c$, $T_{in}^h$) and outlets ($T_{out}^c$, $T_{out}^h$) of both the cold and hot fluids. The volumetric flow rates of the hot and cold fluids, denoted by $\dot{V}^h$ and $\dot{V}^c$, respectively, are also recorded. Heat exchange is regulated using a motorized valve, with the control input $\theta^c$.}
\label{fig:Gipsa_Heat_Diagram}
\end{figure*}

This study focuses on the counter-flow configuration, in which the hot and cold fluids flow in opposite directions. Fig.~\ref{fig:Gipsa_Heat_Diagram} shows a schematic representation of the setup, including the temperature and flow sensors and the actuated valve.
The hot fluid in the heated water tank is maintained at $T_{in}^h = 60$\degree{}C using a thermostat. It is circulated by a pump, and its volumetric flow rate remains constant (PI-regulated) during the experiments. The flow rate is regulated by an actuated valve based on the measurement provided by the volumetric flow meter, $\dot{V}^{h}$. The heat from the hot fluid is transferred to the cold fluid in the HE, after which the hot fluid flows directly back to the hot tank.
The cold fluid follows an independent path in a similar HE-tank circuit. It flows in the opposite direction through the HE, and its volumetric flow rate, $\dot{V}^{c}$, is the regulated variable used to control the heat exchange. The temperature of the cold tank is not regulated.


\subsection{Overview of the Control Architecture}

\begin{figure*}
\centering
\includegraphics[width=14cm]{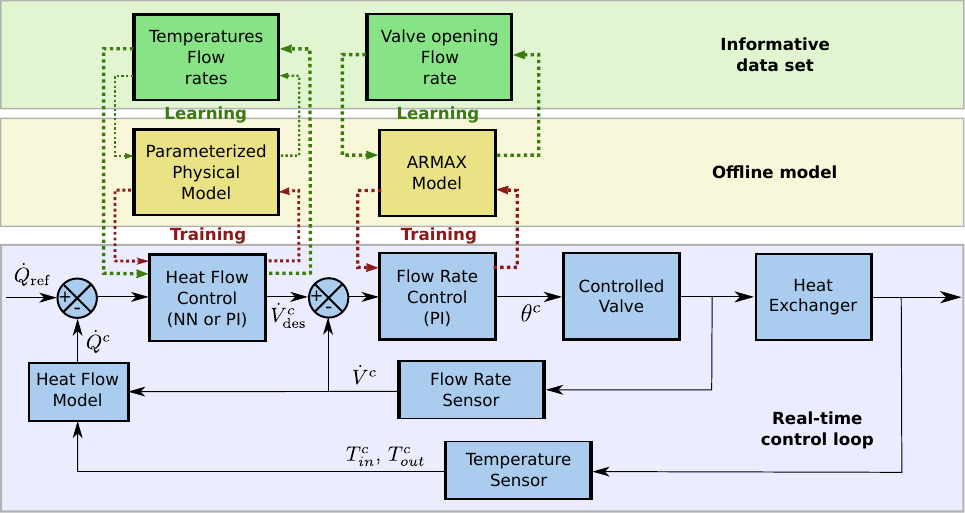}
\caption{Control system architecture, including the learning process and real-time feedback loops}
\label{fig:control_schematic}
\end{figure*}

The proposed control architecture, shown in Fig.~\ref{fig:control_schematic}, includes two nested control loops: an \textit{inner loop} for volumetric flow-rate control and an \textit{outer loop} for heat-flow-rate regulation.
Each control loop includes a reference data set, an offline model, and a real-time controller. The models are designed to capture the time scales relevant to each loop: fast dynamics in the inner loop, dictated by the response time of the actuated valve and the fluid flow, and slower dynamics in the outer loop, governed by the thermal behavior of the HE.
Each model depends on tunable parameters that are identified through a learning process based on the reference data set. This data set must be informative, with sufficient amplitude and frequency content to accurately represent the desired operating conditions and distinguish between candidate models.
The real-time controllers are trained using the offline models, possibly supplemented by an additional experimental data set, to achieve precise tuning. This model-based training significantly reduces the amount of experimental time required compared with black-box approaches, which do not rely on a physical model, or online methods such as reinforcement learning, which require many iterative closed-loop experiments with different control signals to obtain an effective control policy.
Moreover, the physical models can be used in real time as estimators, providing useful information for system diagnostics and predictive maintenance.

\section{Valve Modeling and Volumetric Flow Rate Regulation}
\label{sec:innerloop}

Accurate volumetric-flow-rate regulation is required to achieve effective heat-flow-rate control. This section describes how the inner control loop accomplishes this objective by controlling the motorized proportional valve on the cold-fluid circuit using the measured volumetric flow rate, $\dot{V}^c$. The water pump operates continuously at a fixed manometric head, while the volumetric flow rate is adjusted through the actuated valve.

\subsection{Experiments for Dynamic Modeling}

\begin{figure}
\centering
\includegraphics[width = \linewidth]{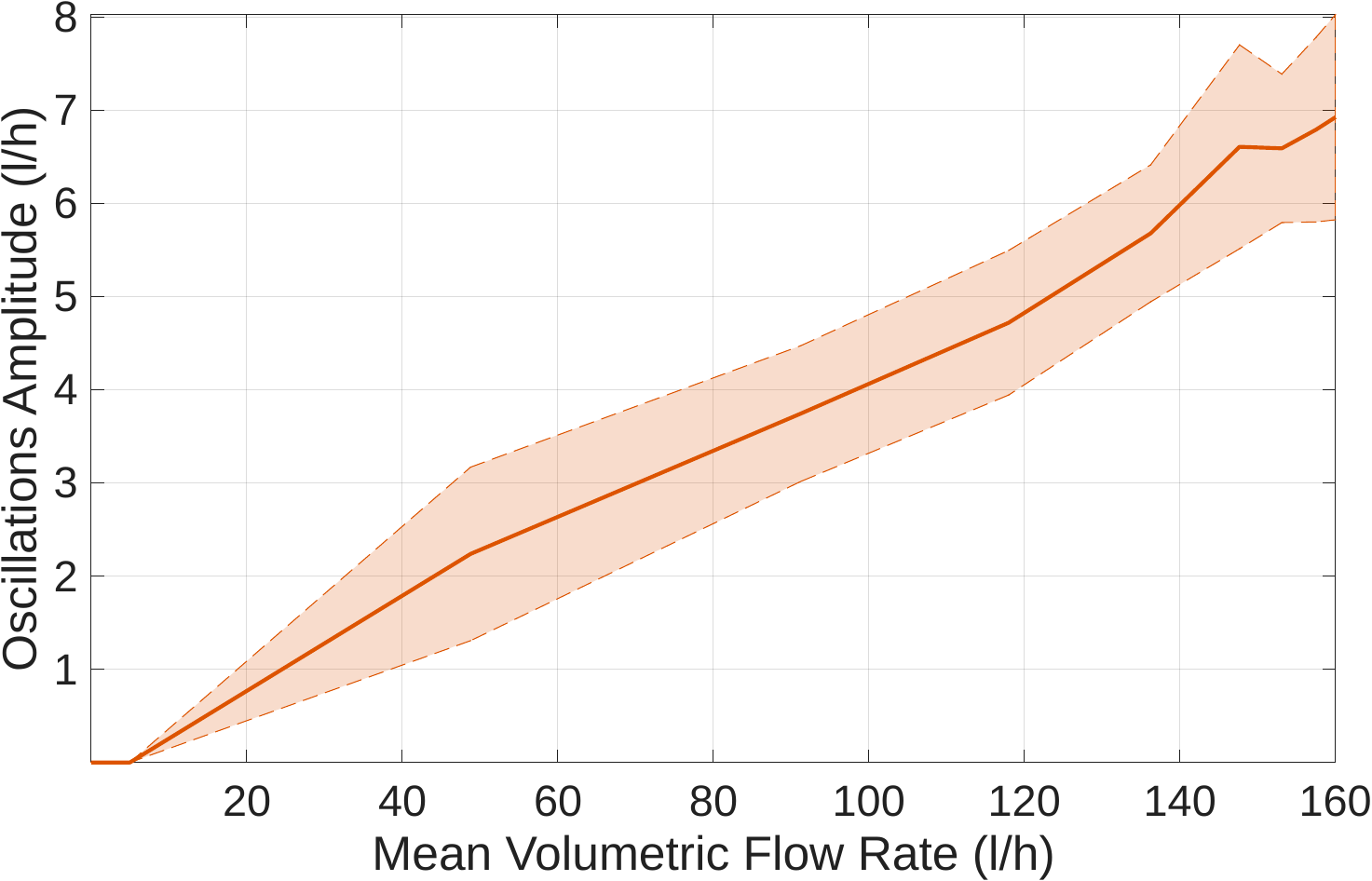}
\caption{Mean amplitude and standard deviation of the volumetric flow oscillations measured when modulating the volumetric flow rate with the regulated valve.}
\label{fig:Noise Amplitude}
\end{figure}

Preliminary experiments on the motorized valve, described in Appendix~\ref{App:PreExp}, show that the static gain of the valve varies significantly with $\dot{V}^c$. The volumetric flow rate also affects the amplitude of the high-frequency oscillations, shown in Fig.~\ref{fig:Noise Amplitude}. These two nonlinear effects can strongly influence control performance.
The control design therefore requires a reference data set that accurately captures the frequency response over the operating range of interest, providing a relatively high-order dynamic model suitable for precise tuning of a proportional–integral (PI) controller.

\begin{figure}
\centering
\includegraphics[width=\linewidth]{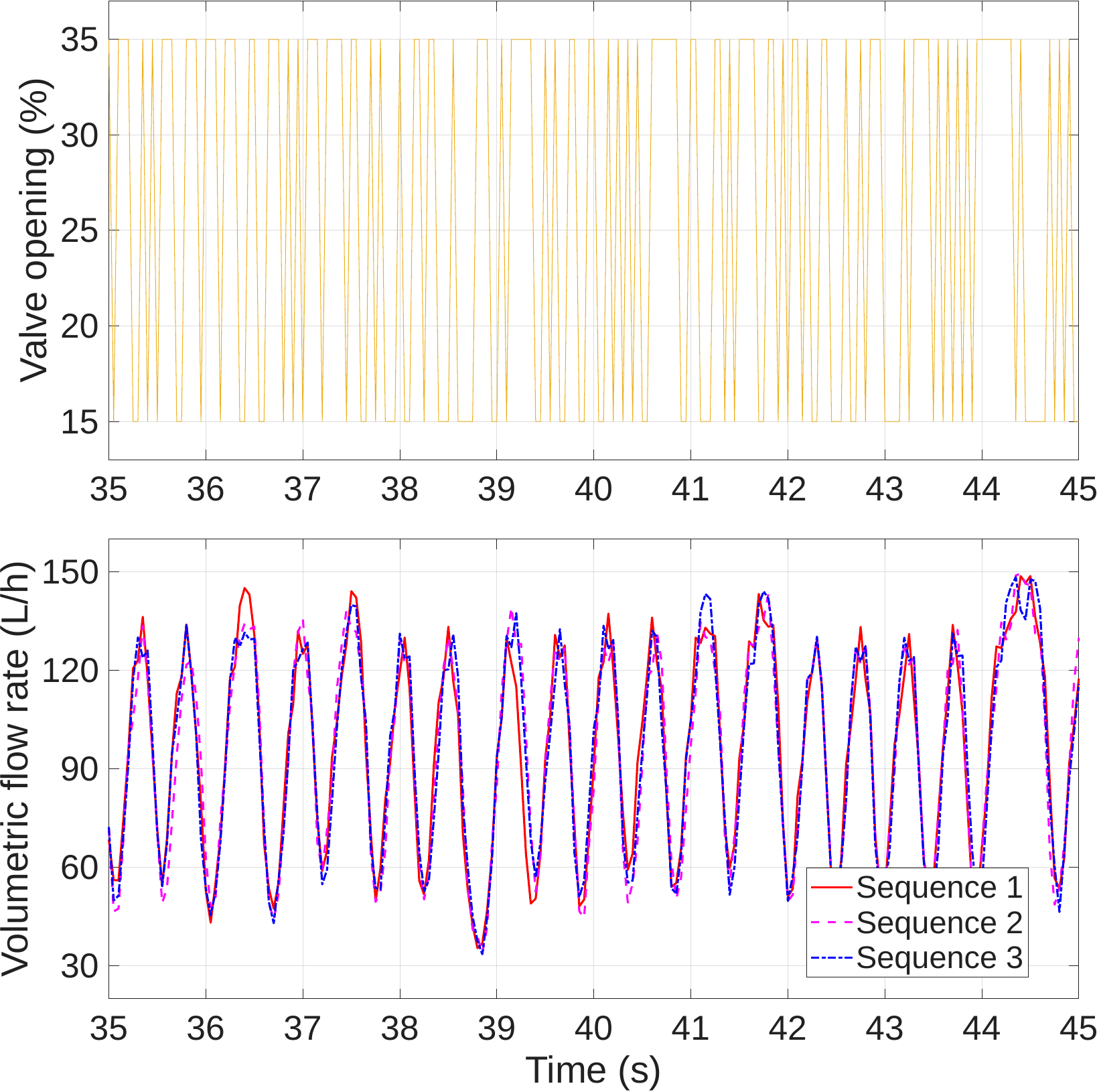}
\caption{Flow response to a PRBS input signal applied to the motorized valve (top) and corresponding responses over three repetitions (bottom); zoom over a 10\,s time window.}
\label{fig:PRBS}
\end{figure}

A sampling time of $t_s=50~\mathrm{ms}$ is selected based on preliminary valve-identification experiments, and an input signal that is sufficiently rich to excite all relevant frequencies is chosen. Specifically, a tenth-order pseudo-random binary sequence (PRBS), designed according to the method described by \cite{10.5555/21413}, is applied to the valve.
This signal is centered at 25\,\% of the input range with an amplitude of 10\,\%. The resulting input range, spanning from 15\,\% to 35\,\%, is expected to produce a volumetric flow rate between 30 and 140\,L/h, thereby covering a broad operating range. The duration of the PRBS is given by
$
(2^N-1)t_s \approx 51~\mathrm{s},
$ 
corresponding to 1023 samples with $N=10$. The input sequence and the corresponding system response over three consecutive cycles are shown in Fig.~\ref{fig:PRBS}. A high degree of repeatability can be observed among the experimental responses.

\subsection{Method: Offline Model and Controller Design}

A discrete-time transfer function is estimated from the PRBS input–output data by identifying a discrete-time state-space model using the subspace method \citep[][implemented in MATLAB\textsuperscript{\textregistered}'s \texttt{n4sid} function]{10.5555/21413}.
More specifically, the method generates a Hankel singular-value plot representing the relative energy contribution of each system state. An appropriate model order is then selected based on the dominant singular values, and the corresponding state-space model is identified.
For the valve, this method yields a fourth-order model with a one-sample delay, induced by the zero-order hold used during data acquisition. The resulting discrete-time transfer function (ARMAX model) is
\begin{equation}
\frac{\dot{V}^c}{\theta^c} (z^{-1}) = 
\frac{6.48  \, z^{-1} + 13.11 \, z^{-2} - 7.87 \, z^{-3} + 2.71 \, z^{-4}}
{1 - 2.75 \, z^{-1} + 3.05 \, z^{-2} - 1.51 \, z^{-3} + 0.25 \, z^{-4}},
\label{eq:Discrete-time Transfer Function}
\end{equation}
where $z$ denotes the $\mathcal{Z}$-transform operator and $\theta^c$ is the control signal (expressed in \% of the input range) applied to the valve actuator.

Based on the identified transfer function in Eq.~\eqref{eq:Discrete-time Transfer Function}, a PI controller is designed to ensure accurate tracking of the desired volumetric flow rate. The MATLAB\textsuperscript{\textregistered} PID Tuner is used to determine suitable controller gains. This tool enables interactive tuning while accounting for both performance and robustness requirements of the closed-loop system.
Starting from the identified transfer function, the user can specify desired characteristics such as response time, bandwidth, transient behavior, and phase margin. The tuning algorithm then automatically adjusts the controller gains to satisfy the prescribed performance and robustness objectives.
The primary design objective is to achieve a rise time of approximately 1\,s while maintaining accurate reference tracking. The resulting PI gains are
$K_P = 2.36\times10^{-4}$ and $K_I = 5.24\times10^{-3}$.
The resulting closed-loop response is close to that of a slightly underdamped second-order system, which is generally regarded as a good compromise between performance and robustness \citep{Landau2006}.

\subsection{Results and Discussion}

\begin{figure}
\centering
\includegraphics[width=\linewidth]{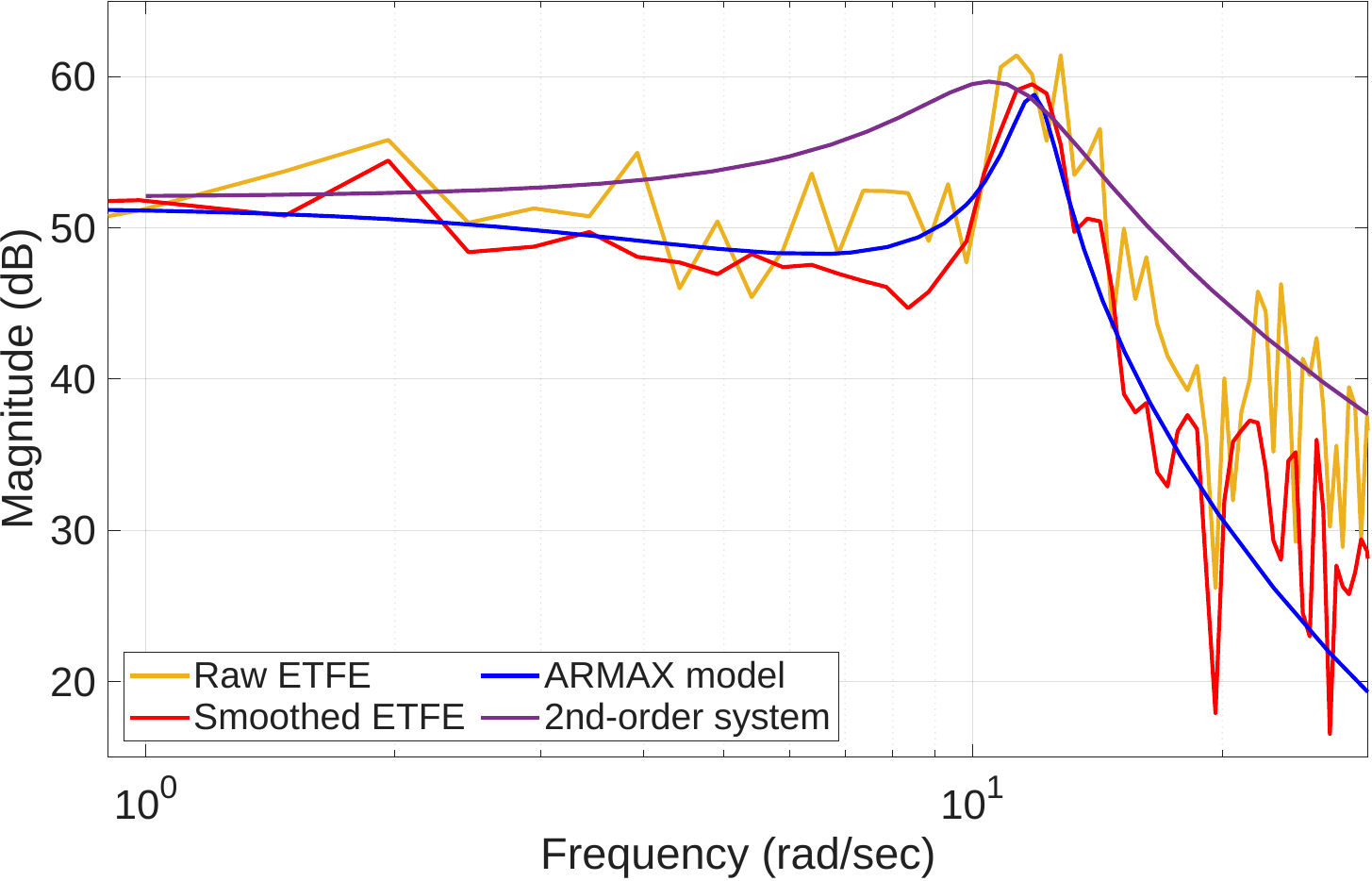}
\caption{Valve model evaluation: ETFE (raw and smoothed) of the measured sequences of $\dot{V}^c$ resulting from the PRBS input, and frequency responses of the identified ARMAX model and of a 2$^{nd}$ order system.}
\label{fig:ETFE Response}
\end{figure}

The ability of Eq.~\eqref{eq:Discrete-time Transfer Function} to reproduce the transient behavior of the valve is assessed through frequency-domain analysis.
The frequency response of the experimental system is obtained by computing the empirical transfer function estimate (ETFE) using data from the three experiments (3069 samples). The ETFE is subsequently smoothed using a Hamming window of length 150, following the guidelines of \cite{10.5555/21413} (choosing a length equal to the number of samples / 20). This smoothing operation reduces variance while preserving the underlying causal dynamics of the physical system.
The raw ETFE, the smoothed ETFE, and the frequency response of the identified transfer function in Eq.~\eqref{eq:Discrete-time Transfer Function} are shown in Fig.~\ref{fig:ETFE Response}. For comparison, a classical second-order model with a natural frequency of 11\,rad/s and a single zero, often used to represent electronic valves \citep{Johnson1965,WITRANT2023105634}, is also displayed.
While the second-order model fails to capture both the sharp resonance peak around 10\,rad/s and the high-frequency roll-off, the ARMAX model of Eq.~\eqref{eq:Discrete-time Transfer Function} provides a much closer approximation of the experimentally observed dynamics.

\begin{figure}
\centering
\includegraphics[width = \linewidth]{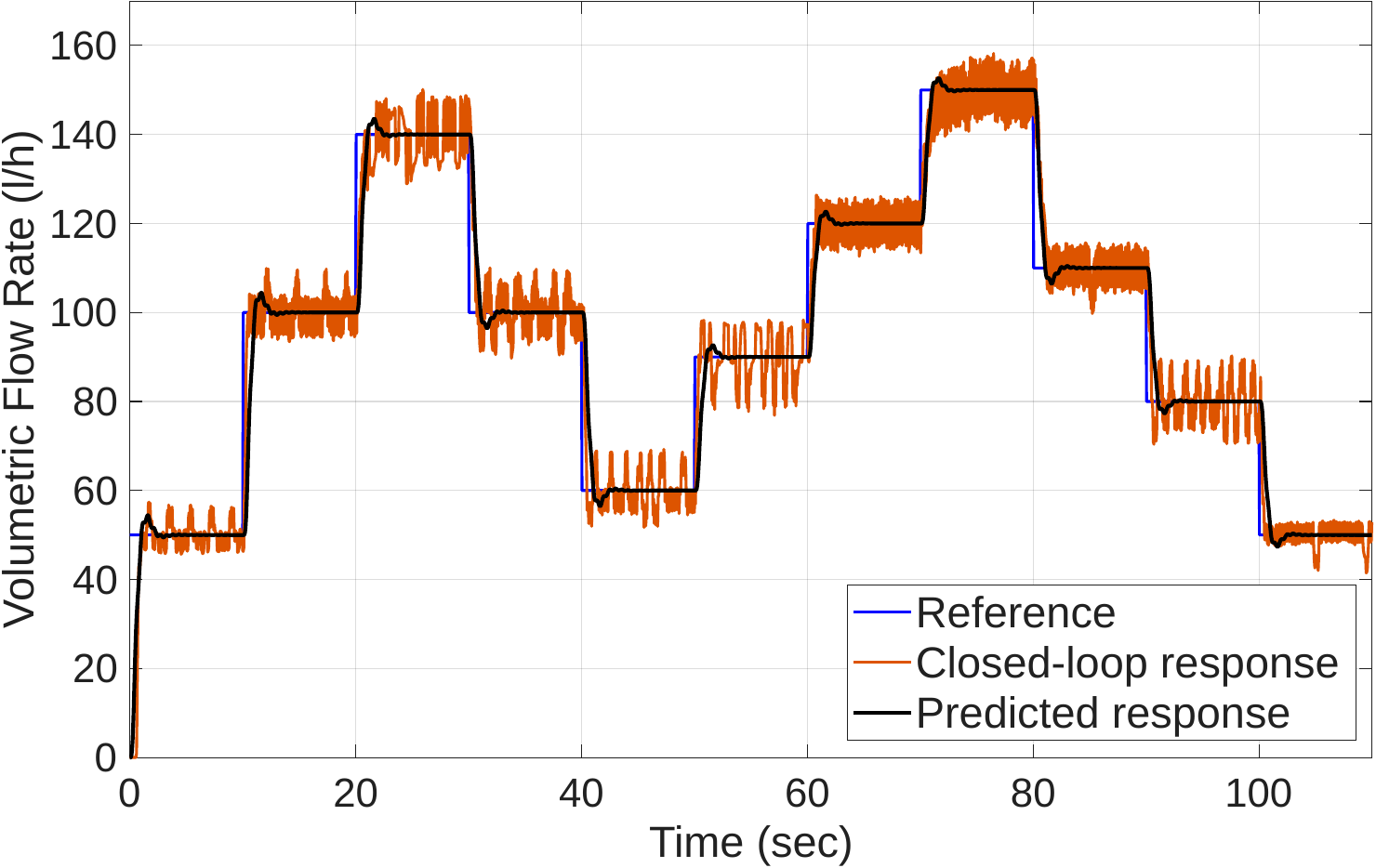}
\caption{Closed-loop regulation of $\dot{V}^c$: measured and predicted capabilities to track an arbitrary step-wise reference signal.}
\label{fig:Closed-loop valve responses to ref}
\end{figure}

The real-time closed-loop response obtained with the proposed PI controller is shown in Fig.~\ref{fig:Closed-loop valve responses to ref}. The figure also includes the response predicted during the tuning stage.
These results demonstrate the controller's ability to track the reference signal over the considered operating range. Although the controller is designed from a local linear model, the average closed-loop response remains in close agreement with the desired behavior throughout the operating domain.
The high-frequency oscillations observed in the measured flow rate, similar to those reported in the open-loop experiments of Fig.~\ref{fig:Noise Amplitude}, cannot be fully compensated by such a simple controller. Their effect on heat-flow-rate regulation is addressed by the more advanced control strategy presented in Section~\ref{sec:outerloop}.

\section{Dynamic Model of Heat Transfer and Transport Phenomena} \label{sec:physical_model}

A physics-based model, tailored to the experimental test bench using a dedicated data set, is proposed in this section.
This model must capture heat transport phenomena and nonlinearities to cover a wide range of operating conditions while remaining computationally efficient for parameter optimization.
The dynamics are described by a classical one-dimensional transport equation derived from energy conservation principles and tailored to the specific HE by parameterizing the heat-transfer coefficient and heat losses.
The outer-loop controller relies on this model for both tuning and the first-stage training of the NN used in the feedback law.

\subsection{Reference Data Set for Dynamic Modeling} \label{sec:h_exp}

\begin{figure}
\centering
\includegraphics[width=1.01\linewidth]{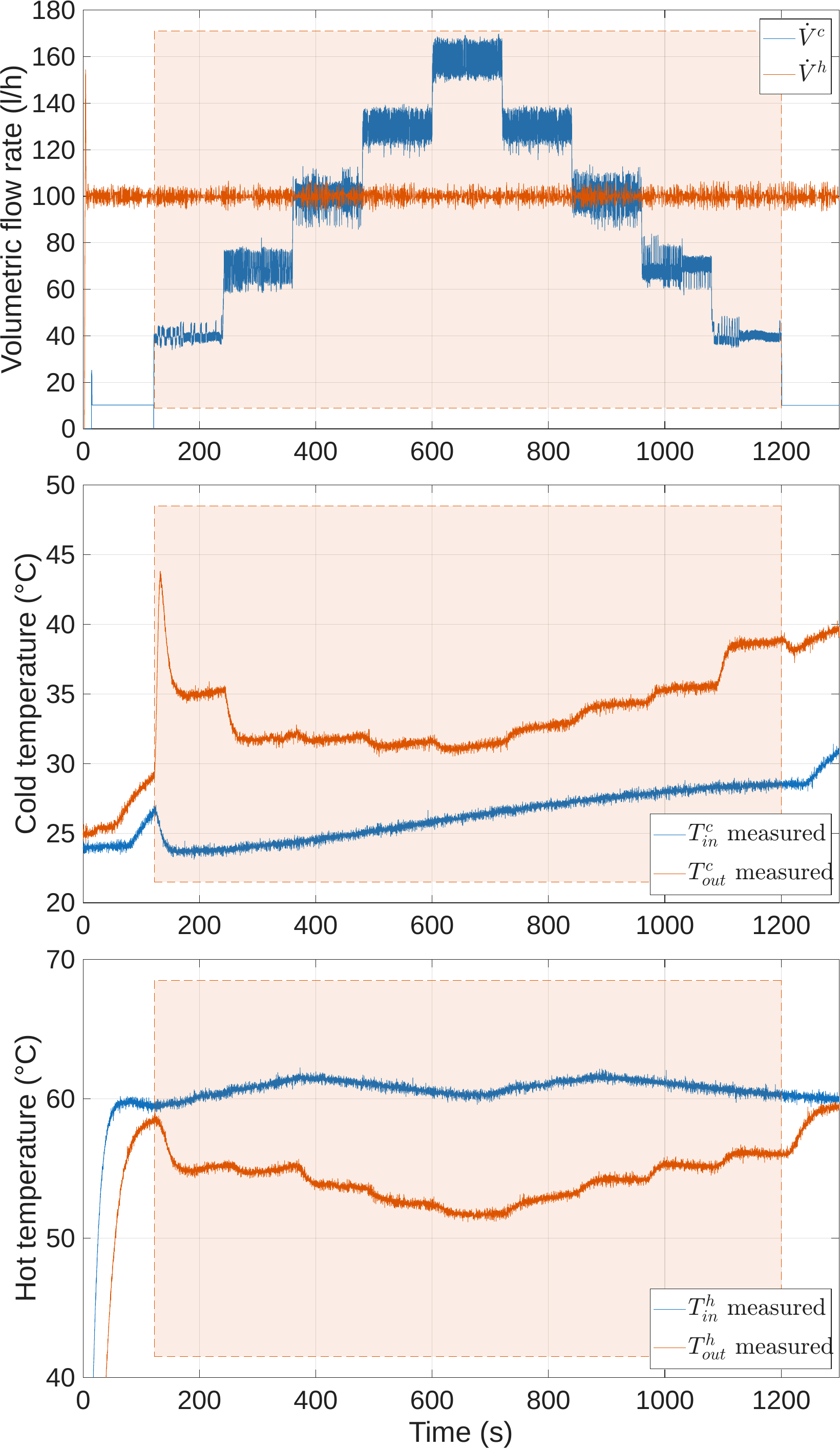}
\caption{Experimental data used to parameterize the heat-transfer coefficient (only the shaded orange area is selected to identify the parameters).
Top: modulations of the cold flow rate and constant hot flow rate.
Middle: evolution of the cold flow inlet and outlet temperatures.
Bottom: evolution of the hot flow inlet and outlet temperatures.
}
\label{fig:h estimation experiment} 
\end{figure}

The data set used to determine the free coefficients of the model is presented in Fig.~\ref{fig:h estimation experiment}.
The volumetric flow rate of the hot fluid is kept constant at 100\,L/h, and a gradually increasing and decreasing sequence of step functions from 0 to 160\,L/h is applied to $\dot{V}^c$, as depicted in the top panel.
The resulting hot and cold temperatures (inlets and outlets) are measured and presented in the middle and bottom panels, respectively.
The ambient temperature is 23.8\degree{}C.
A discrepancy in the signals can be observed at low $\dot{V}^c$ values (around 10\,L/h), where the water circuit is started at room temperature and specific transient phenomena reflect the time required for the hot-water circuit to homogenize with the hot-water tank.
As these transient effects are not included in the proposed model and the regulation objectives correspond to higher volumetric flow rates, the reference data set is restricted to flow rates for which $\dot{V}^c$ is higher than 40\,L/h (orange area on the figure).
Furthermore, the high-frequency components of the signals are removed by applying a standard trend filter (8$^{\textrm{th}}$-order digital moving-average filter).

\begin{figure}
\centering
\includegraphics[width=\linewidth]{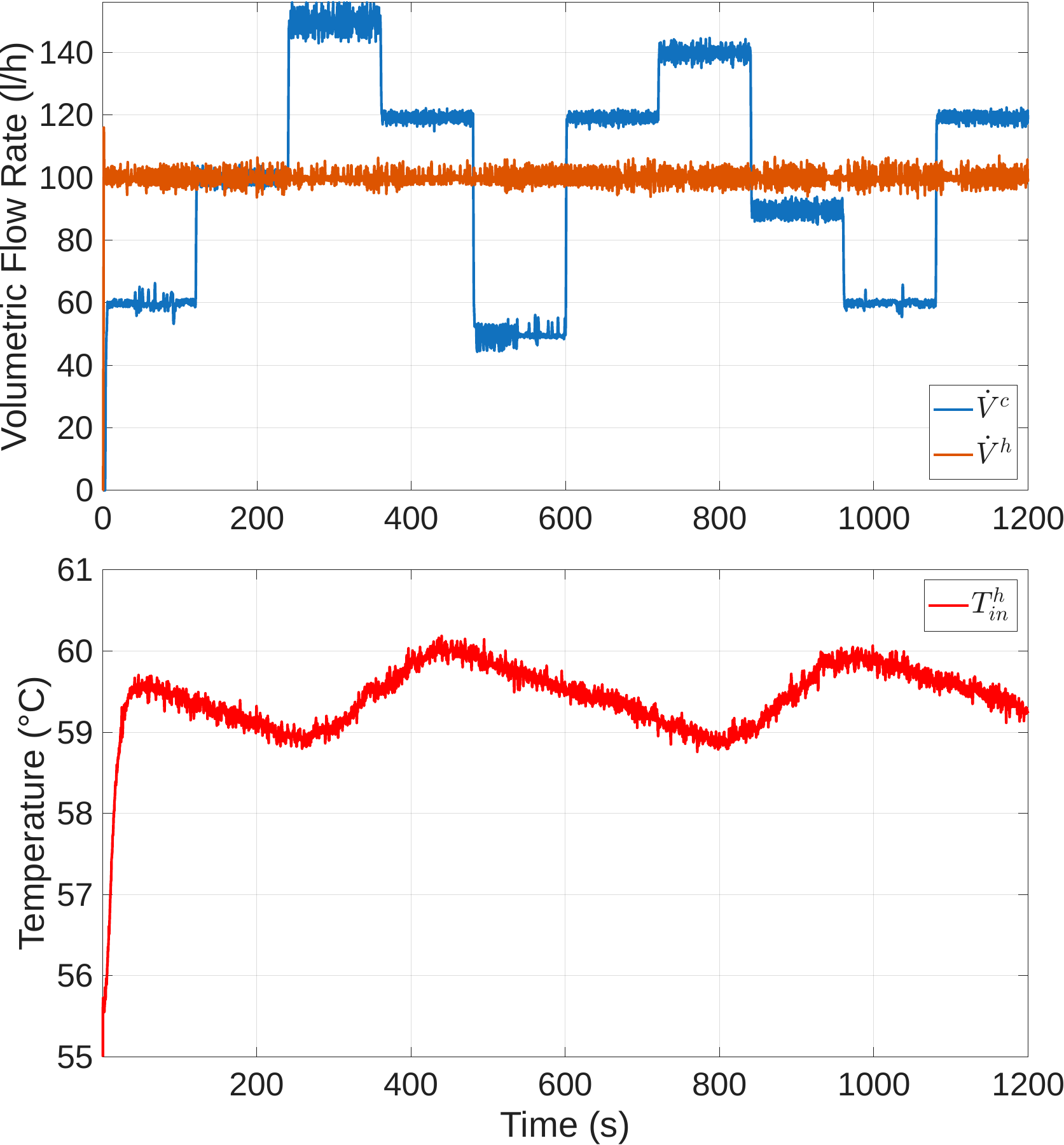}
\caption{Evaluation of the physical model: hot and cold volumetric flow rates (top), and hot temperature input (bottom).}
\label{fig:Simulator_valid_InFlows}
\end{figure}

To evaluate the predictive capability of the model on an independent data set, a second experiment is designed with a constant $\dot{V}^h$ and modulated values of $\dot{V}^c$, as depicted in Fig.~\ref{fig:Simulator_valid_InFlows}.
Compared with the experiment used for parameter estimation, steps of different (larger) amplitudes are generated on $\dot{V}^c$, and the ambient temperature is significantly higher (32\degree{}C).
For both data sets, the heater in the hot tank partially compensates for the heat exchange, and slow variations can be observed in $T^h_{in}$.
These variations are included as known boundary conditions of the model.

\subsection{Methods}

The schematic representation of the simplified HE model is illustrated in Fig.~\ref{fig:Gipsa_Heat_Diagram} (top-middle part).
It consists of two volumes of length $L$, separated by a wall.
In the outer volume, a hot fluid with temperature $T^h(x,t)$ flows rightward.
In the inner volume, a cold fluid with temperature $T^c(x,t)$ flows in the opposite direction.
To analyze the temperature dynamics of the two flows within the HE, several classical simplifying assumptions are introduced \citep{bonne:tel-01133432}:
(i) the flow is one-dimensional (along an axis denoted by $x$ in Fig.~\ref{fig:Gipsa_Heat_Diagram}), with the hot fluid moving in the positive direction;
(ii) heat diffusion within the inner and outer tubes is negligible;
(iii) the wall thickness is neglected, and the heat-transfer coefficient is assumed to be uniform and constant in space;
(iv) the connecting pipes and the cold tank are perfectly insulated and do not exchange heat with the surrounding environment;
(v) densities and specific heats are constant over the temperature range considered in the experiments.
Under these assumptions, heat exchange occurs solely between the two fluids through the inner HE tube and between the hot fluid and the ambient air through the outer HE tube.
The only relevant mechanism considered in the energy conservation model is convective heat transfer.

\subsubsection{Heat transport model}

The impact of volumetric-flow variations on the spatial temperature field is included by parameterizing their influence on the heat-transfer coefficient.
The transport equations used in the physical model are written as follows:
\begin{align}
E^h \pdv{T^h}{t} + F^h \pdv{T^h}{x} &= - \frac{1}{L} \dot{Q}^h  \label{eq:PDE_Th}
 \\
E^c \pdv{T^{c}}{t} - F^c \pdv{T^{c}}{x} &= h^c p \big(T^{h}-T^{c}\big)
\label{eq:PDE_Tc}
\end{align}
where 
$$E^{h,c} = \rho^{w} C_p^w S^{h,c} + \frac{m^{inox} C_p^{inox}}{2 L},$$ $F^{h,c} = \dot{V}^{h,c}(t) \rho^{w} C_p^w$,
$C_p^{w}$ denotes the specific heat of water at constant pressure, $\rho^w$ the water density, $h^{c}(t)$ is the heat-transfer coefficient related to the cold fluid convection (defined in the next subsection), $p$ is the exchange perimeter in the HE cross-section, $S^{h,c}$ are the cross-sectional areas of the hot and cold fluid, respectively, $m^{inox}$ and $C_p^{inox}$ are the mass and specific heat of 316L stainless steel, and $\dot{Q}^h$ is the heat flow rate lost by the hot fluid.
The system of Eqs.~\eqref{eq:PDE_Th}-\eqref{eq:PDE_Tc} has known boundary conditions:
\begin{align}
   T^h(0,t) &= T^h_{in}(t)    \label{eq:BC_Th}  \\
   T^c(L,t) &= T^c_{in}(t)   \label{eq:BC_Tc}
\end{align}
and initial conditions:
\begin{align}
   T^h(x,0) &= T^h_0(x)     \label{eq:IC_Th} \\
   T^c(x,0) &= T^c_0(x)  \label{eq:IC_Tc}
\end{align}

The cold outlet of the HE is connected to its cold inlet through a tank of volume $V^{tank}$ and temperature ${T}^{tank}$. Neglecting heat losses during water transport and in the tank, the flow entering the tank has a temperature $T_{out}^c$ while the flow leaving the tank sets the temperature $T_{in}^c$.
Conservation of energy sets a relationship between these two variables as first-order dynamics:
\begin{align}
\dot{T}^{tank} &= \frac{\dot{V}^c}{V^{tank}}(T_{out}^c - {T}^{tank}) \label{eq:Ttank} \\
{T}_{in}^c &= {T}^{tank} \label{eq:Ttank_out}
\end{align}
and ${T}_{in}^c(0) = T^{tank}_0$.

For the numerical implementation of this model, the spatial derivatives in the system of PDEs \eqref{eq:PDE_Th}-\eqref{eq:PDE_Tc} are discretized using a backward (upwind) finite-difference scheme for the hot flow and a forward one for the cold flow.
The time evolution of the resulting ordinary differential equations (ODEs) is computed by delaying the nonlinear component of the equations by one sample and then applying an implicit time discretization, as proposed by~\cite{Witrant_2007}.
This method is particularly efficient for obtaining numerical models of PDEs that operate in real time, provided that the nonlinearities are sufficiently smooth, which is the case here.

\subsubsection{Heat transfer coefficient}

The physical model is tailored to the experimental test bench by parameterizing the heat-transfer coefficient $h$ according to the reference data set presented in Section~\ref{sec:h_exp}.
This parameter is crucial because it governs the amount of heat energy transferred from the hot side to the cold side.
It depends on various factors, including the type of medium (gas or liquid), the flow velocity, and the physical properties of the fluid, such as viscosity and other temperature-dependent quantities.

Consequently, the behavior of $h$ is highly complex and can exhibit significant variations in response to small changes in these influencing parameters.
An accurate determination of $h$ is generally possible only under steady-state conditions and requires an exhaustive experimental procedure grounded in fluid-mechanics principles.
This typically involves the evaluation of the Prandtl (Pr), Reynolds (Re), Nusselt (Nu), or Stanton numbers \citep{Gnielinski75, VERAGARCIA20101231}.
This problem is simplified here as only single-phase fluids are considered (constant Pr).
In that case the log-mean temperature difference (LMTD) method under steady-state conditions used by \cite{Gnielinski75} has been shown to provide an accurate estimation.
As energy conservation is easier to express for the cold fluid (which gains all its heat from the hot fluid) than for the hot fluid (which also dissipates heat to the surrounding environment), the cold fluid is chosen for determining the heat-transfer coefficient.

The convective heat-transfer coefficient estimated with the LMTD method $h_{LM}$ can be calculated in steady-state conditions using:
\begin{equation} 
 \dot{Q}^c_{SS} = h_{LM} S^{he}_{in} \Delta{T_{LM}}
  \label{eq:LMTD7}
\end{equation}
where $\dot{Q}^c_{SS}$ is the heat flow rate received by the cold fluid, 
$S^{he}_{in}=p L$ the heat exchange surface of the inner tube, and
\begin{equation}
\Delta{T_{LM}} \doteq \frac{(T_{in}^h - T_{out}^c) - (T_{out}^h - T_{in}^c)}{\ln{[(T_{in}^h - T_{out}^c)/(T_{out}^h - T_{in}^c)}] }
 \label{eq:LMTD8}
\end{equation}
is the LMTD averaging the temperature difference along the inner tube.
The convective heat transfer is related to the energy transported by flow advection by integrating Eq.~\eqref{eq:PDE_Tc} over space at steady-state, which provides the following relationship:
\begin{equation}
 \dot{Q}^c_{SS} = \dot{V}^c \rho^{w} C_p^w (T_{out}^c-T_{in}^c )  \label{eq:LMTD-Heat Rate Cold}
\end{equation}
Combining Eqs.~\eqref{eq:LMTD7} and \eqref{eq:LMTD-Heat Rate Cold}, $h_{LM}$ is calculated from experimental data as:
\begin{equation}
h_{LM} = \frac{\dot{V}^{c} C_p^w \rho^{w} (T_{out}^c-T_{in}^c )}{S^{he}_{in} \Delta{T_{LM}}} \label{eq:h_calculation}
\end{equation}

Using the reference value of $h_{LM}$ calculated from the data set with the previous formula, $h^c$ is parameterized using a Dittus--Boelter correlation \citep{Dittus1930} that defines the relationship between the Nusselt and Reynolds numbers.
This correlation has been shown to be particularly efficient for a wide range of operating conditions (e.g. see \cite{TALER2016108} and references therein).
As transitions from laminar to turbulent flows are observed in the data set described in Section~\ref{sec:h_exp}, the two-regimes formula (for Re < 2300 and Re $\geq$ 2300) proposed by \cite{GNIELINSKI2013134} is used here. 
The transition between the regimes, discontinuous in \cite{GNIELINSKI2013134}, is rendered continuous here using a sigmoid function $\gamma(\textrm{Re})$ to provide a more robust numerical scheme, and the scaling law reads:
\begin{align}
\textrm{Nu} &= (1 - \gamma) a_{lam,1} \textrm{Re}^{a_{lam,2}} \textrm{Pr}^{0.4} \nonumber \\
&\quad +  \gamma a_{turb,1} \textrm{Re}^{a_{turb,2}} \textrm{Pr}^{0.4} \label{eq:Nu_scaling} \\
\gamma &= \frac{1}{1 + \exp\left(-(\textrm{Re} - 2300)\right)} \label{eq:Nu_scaling_sigma}
\end{align}
where $\textrm{Re} = \Dot{V}^c \rho^{w} d_H / (\mu A_c)$, $\textrm{Pr} = \mu C_p^w /k$, $d_H$ is the hydraulic (inner) diameter of the inner tube, $k$ the thermal conductivity, $\mu$ the water viscosity and $A_c= \pi d_H^2/4$ is the free flow area.
The 0.4 exponent on Pr is chosen by similarity to the classical Dittus--Boelter correlations. 
The constant parameters describing the laminar ($a_{lam,1,2}$) and turbulent ($a_{turb,1,2}$) behaviors are obtained by minimizing the squared error between
$\textrm{Nu} = h_{LM} d_H / k$ and the right-hand side of Eq.~\eqref{eq:Nu_scaling}.
The heat-transfer coefficient in the transport equation \eqref{eq:PDE_Tc} is then calculated as $h^c = \textrm{Nu}\, k / d_H$ for a given Re (thus $\dot{V}^c$) using the scaling Eqs.~\eqref{eq:Nu_scaling}-\eqref{eq:Nu_scaling_sigma}.

Ideally, for a perfectly insulated HE, the heat flow rate lost by the hot fluid $\dot{Q}^h$ is equal to the heat flow rate gained by the cold fluid $\dot{Q}^c$. 
However, heat losses are observed in the experiment studied in this paper and the energy dissipation is included as follows.
The heat flow rate lost by the hot fluid in Eq.~\eqref{eq:PDE_Th} is set according to: 
\begin{equation}
\dot{Q}^h = h^c p L \big(T^{h}-T^{c}\big) + \dot{Q}^a \label{eq:dQh}
\end{equation}
where the first term on the right-hand side corresponds to the energy gained by the cold fluid and $\dot{Q}^a$ denotes the heat flow rate dissipated to the environment at ambient temperature $T^a$.
Considering the average temperature difference between ambient air and the hot flow, calculated by the LMTD:
\begin{equation}
\Delta T_{LM}^{a} \doteq \frac{(T_{in}^h - T^a) - (T_{out}^h - T^a)}{\ln{[(T_{in}^h - T^a)/(T_{out}^h - T^a)}] }
 \label{eq:LMTD_ha}
\end{equation}
to quantify the temperature difference between $T^h$ and $T^a$ on the outer tube and a convective heat transfer, one can define the mean (for the reference data set) heat-transfer coefficient between the hot fluid and ambient air as:
\begin{equation}
h_{LM}^a = \textrm{mean} \left(\frac{\dot{Q}^h_{SS} - \dot{Q}^c_{SS}}{S^{he}_{out} \, \Delta T_{LM}^a }  \right)
\end{equation}
where the mean is carried over the steady-state values, $\dot{Q}^c_{SS}$ is obtained from Eq.~\eqref{eq:LMTD-Heat Rate Cold}, $\dot{Q}^h_{SS} = \dot{V}^h \rho^{w} C_p^w (T_{in}^h-T_{out}^h)$, and $S^{he}_{out}$ is the heat exchange surface of the outer tube.
The heat transport Eq.~\eqref{eq:PDE_Th} is thus defined with $\dot{Q}^h$ calculated as in Eq.~\eqref{eq:dQh} and 
\begin{equation}
\dot{Q}^a = h_{LM}^a S^{he}_{out} \, \Delta T_{LM}^a \label{eq:dQa}
\end{equation}

\subsection{Results and Discussion}

\begin{figure}
\centering
\includegraphics[width=\linewidth]{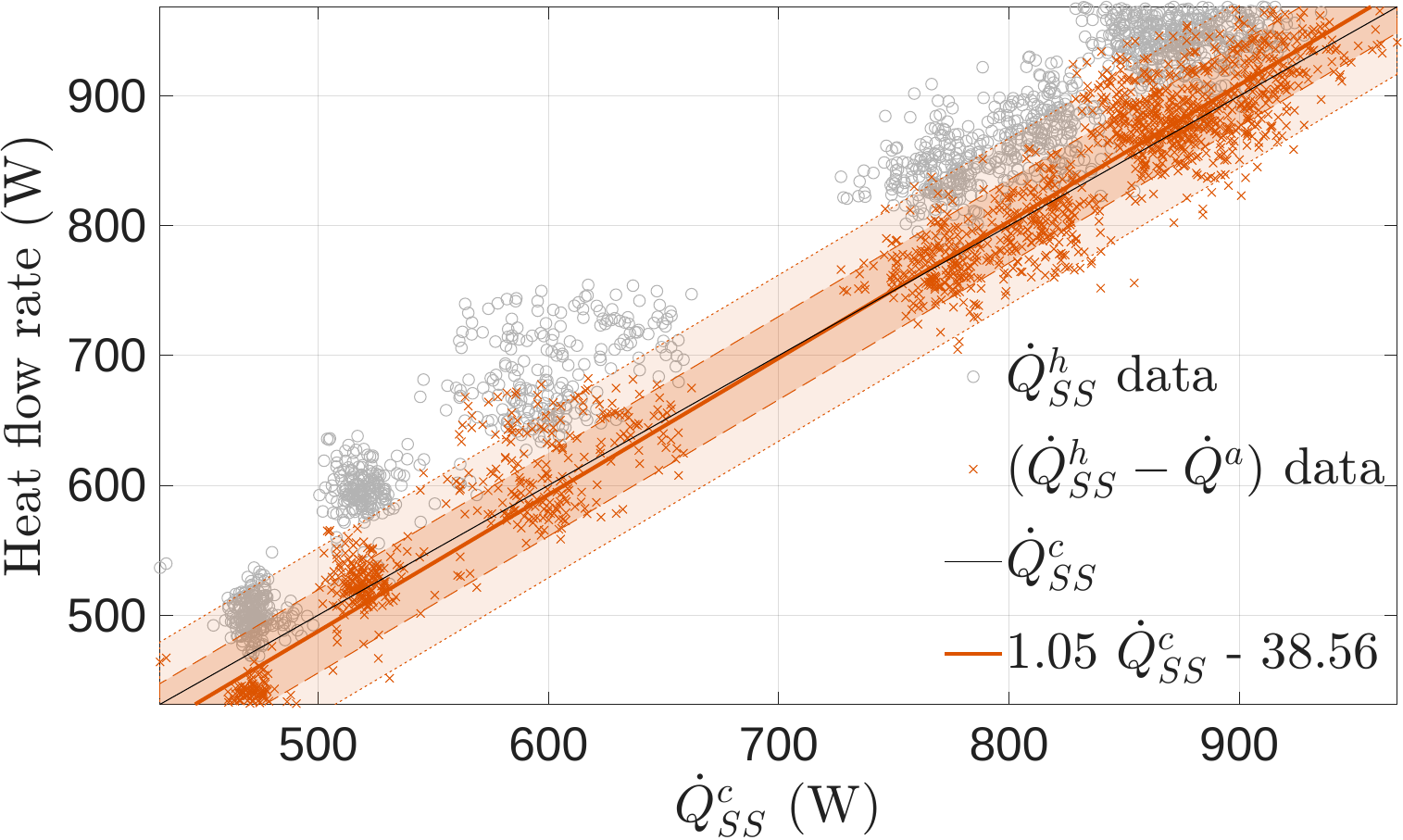}
    \caption{Scaling between the hot and cold heat flow rates at multiple steady-state operating conditions.}
    \label{fig:Qc_vs_Qh_SS}
\end{figure}

The time trends of the heat flow rates are calculated from Eq.~\eqref{eq:LMTD-Heat Rate Cold} (shown in Fig.~\ref{fig:h estimation lmtd} in Appendix~\ref{App:ExtraFigs}).
The steady-state values of the heat flow rates and the accuracy of modeling $\dot{Q}^a$ with Eq.~\eqref{eq:dQa} are shown in Fig.~\ref{fig:Qc_vs_Qh_SS}.
The estimated coefficient is $h^a_{LM} = 24.24$~W/m$^2$.K.
Performing linear regression on $\dot{Q}^h_{SS}-\dot{Q}^a$ (orange line) shows that energy conservation ($\dot{Q}^c_{SS} = \dot{Q}^h_{SS}-\dot{Q}^a$, black line) is verified with a slight deviation (not included in the final model).
Most of the corrected data points lie within the $\pm \sigma$ and $\pm 2 \sigma$ boundaries ($\sigma$ denoting the standard deviation of the estimation error, shaded areas). 

\begin{figure}
\centering
\includegraphics[width=\columnwidth]{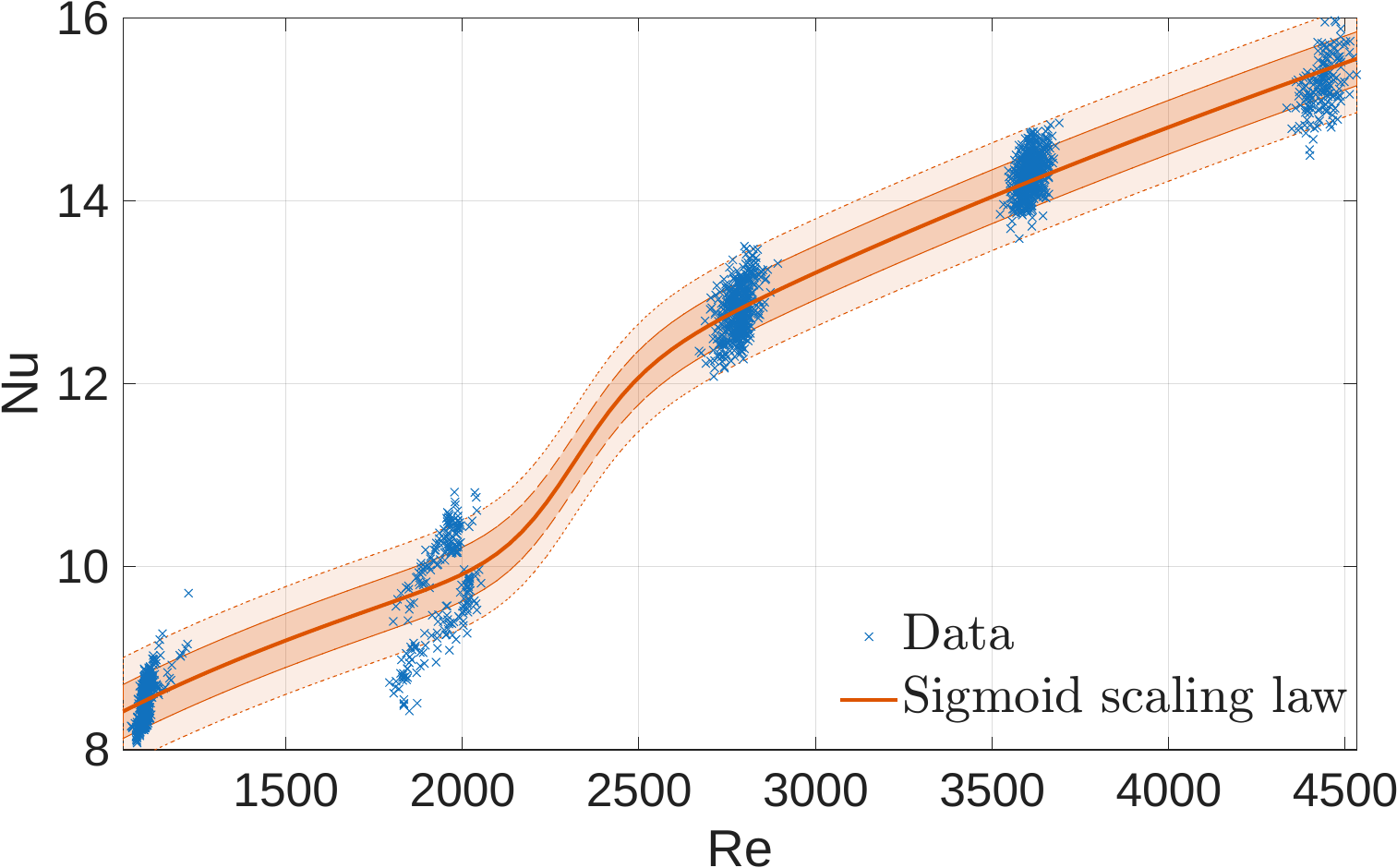}
    \caption{Scaling between Re and Nu numbers used to relate the heat-transfer coefficient to the volumetric flow rate in the heat transport model.}
    \label{fig:Re_vs_Nu}
\end{figure}

The coefficient $h_{LM}$ is computed from Eq.~\eqref{eq:h_calculation} for all the measurements.
The transient data are discarded and only the end of the step responses, once $\dot{Q}^c$ reaches its steady-state value, is kept for the estimation (see Fig.~\ref{fig:h estimation lmtd}, bottom panel).
The same process is repeated several times to cover the whole range of interest for $\Dot{V}^c$.
The resulting data are used as inputs to a least-squares optimization algorithm to compute the coefficients in Eq.~\eqref{eq:Nu_scaling}.
The scaling function in Eqs.~\eqref{eq:Nu_scaling}-\eqref{eq:Nu_scaling_sigma} is depicted in Fig.~\ref{fig:Re_vs_Nu}
and the coefficients are
$a_{lam,1} = 0.726 \pm 0.040$,
$a_{lam,2} = 0.240 \pm 0.007$,
$a_{turb,1} = 0.257 \pm 0.009$, and
$a_{turb,2} = 0.395 \pm 0.004$.
This result shows that the estimated coefficient is in good agreement with the data set for a large range of Re and Nu numbers.
Most of the data are within the $\pm 2 \sigma$ range of the fitting law, except around Re$\approx 2000$, where more fluctuation is observed.
This can be related to the closeness to the laminar-to-turbulent transition, making the flow less predictable.
Comparing the coefficients with those obtained by~\cite{KORZEN2015188}, who calculated $h$ for both laminar and turbulent flow in a tube-and-plate-fin HE, one can notice that in both cases $a_{lam,1} > a_{turb,1}$ and $a_{lam,2} < a_{turb,2}$, with similar ratios between the coefficients.
The accuracy of the laminar-to-turbulent transition could be improved using a friction coefficient parameterized in terms of Re (e.g. see~\cite{ABRAHAM2011584,GNIELINSKI2013134}), but the sigmoid curve in Fig.~\ref{fig:Re_vs_Nu} proved sufficient for the real-time computation and control objectives addressed in this paper.

\begin{figure*}
\includegraphics[width=\linewidth]{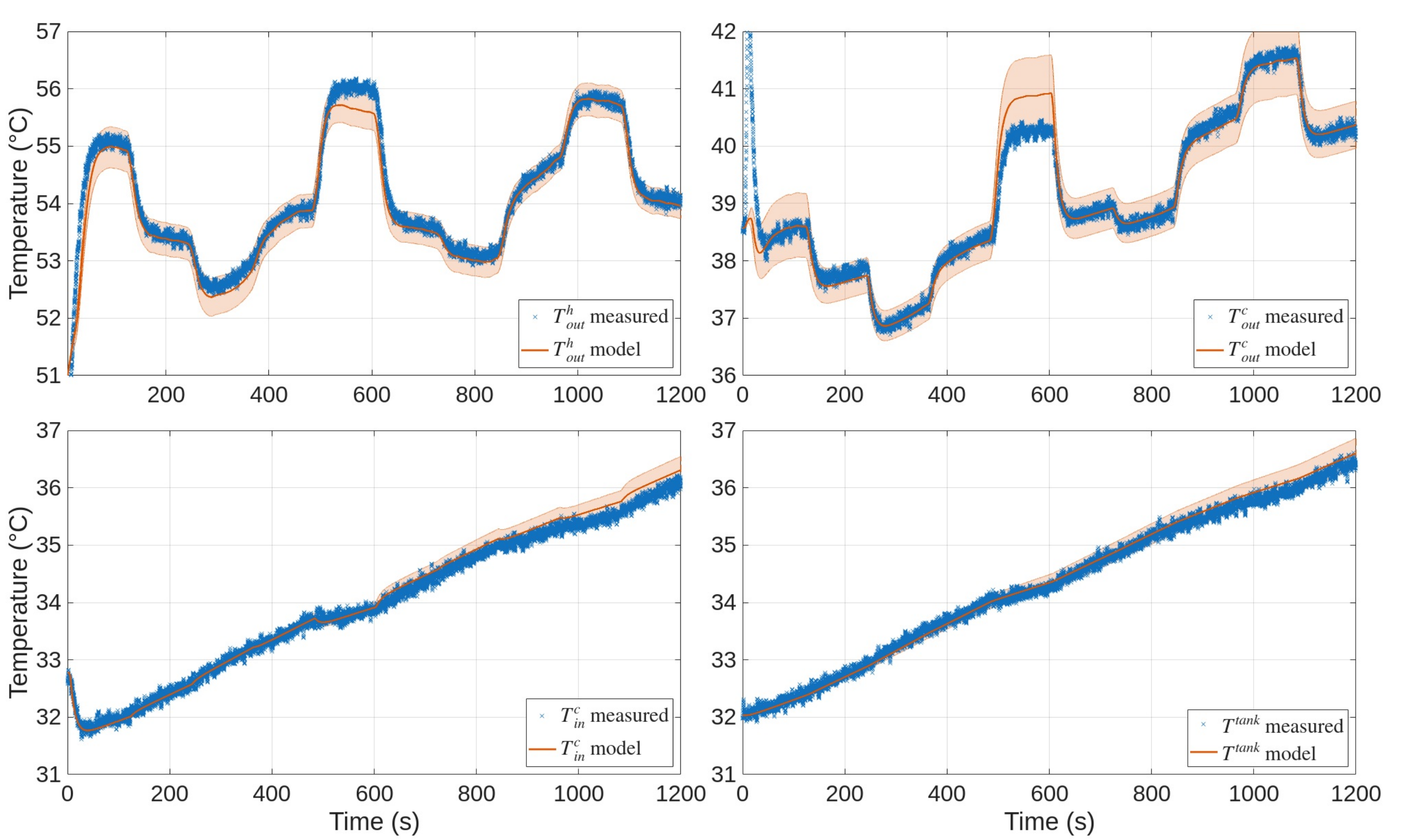}
\caption{Comparison between the physical model temperatures and experimental measurements on the model evaluation data set: hot (top left) and cold (top right) HE outlet temperatures, cold HE inlet temperature (bottom left), and tank temperature (bottom right).}
 \label{fig:Simulator_valid_Out}
\end{figure*}

Combining the estimated coefficient $h^c$ with the physical HE model Eqs.~\eqref{eq:PDE_Th}-\eqref{eq:IC_Tc} and the tank model Eqs.~\eqref{eq:Ttank}-\eqref{eq:Ttank_out} provides a predictive simulator for the HE system. 
This simulator is built with the MATLAB\textsuperscript{\textregistered} software. 
The initial temperatures of the flow in the HE, $T^h_0$ and $T^c_0 = T^{tank}_0$, and the ambient temperature $T^a$ are provided as inputs to the simulator.
The simulated trends are compared with the measurements of $T^h_{out}$, $T^c_{in}$, $T^c_{out}$ and $T^{tank}$ in Fig.~\ref{fig:Simulator_valid_Out}. 
The measured and simulated responses show the high accuracy of the physical model, except between 500 and 600\,s, where a large drop in $\dot{V}^c$ is combined with its lowest value (50\,L/h).
This operating condition corresponds to a low $\dot{Q}^c$ and a less accurate prediction of $\dot{Q}^a$, according to Fig.~\ref{fig:Qc_vs_Qh_SS}.
The simulation results also highlight a slight but progressively increasing difference between the simulated and actual tank temperatures. 
This divergence is due to extra thermal losses to the environment, as the cold tank and connecting pipes are not thermally insulated. 
These losses are not accounted for in the simulation model, as stated in the model assumptions, and could easily be corrected either by following a modeling strategy similar to the one proposed above for the HE, or by introducing a state observer that corrects the state estimate using the temperature sensors in the cold fluid and adaptive algorithms, as proposed by \cite{GHOUSEIN2023111096}.
Overall, the model efficiently captures the transient phenomena and system variations that are essential for control design. 
The computation time of the model is 6.1\,s for the 1200\,s simulation, on a laptop equipped with an Intel\textsuperscript{\textregistered} Core$^{\textrm{TM}}$ Ultra 7 265H × 16 processor.
This fast calculation (about 200 times faster than real-time) makes this model particularly suitable for optimization and real-time estimation and control objectives.

The variations of the estimated heat-transfer coefficient are consistent with changes in flow regime, highlighting the connection between the identified model and classical heat transfer theory.
Increasing the cold flow rate reduces the residence time, leading to faster thermal responses but smaller temperature differences. These observations are consistent with convective transport mechanisms and validate the qualitative physical interpretation of the model.
The experimental results highlight the presence of transport-induced delays between the modulation in $\dot{V}^c$ and the outlet temperatures: while the model accounts for transport within the HE and reacts without additional delay to inlet fluctuations, the measurements (e.g. $T^c_{out}$) tend to react with a short delay.
This is due to the neglected transport time in the connecting tubes and piping, as well as the response time of the temperature sensors (their casing also induces a measurement delay).
This effect has a limited impact here but should be taken into account in larger industrial processes.

\section{Heat Flow Rate Regulation} \label{sec:outerloop}

To track a reference heat flow rate $\dot{Q}_{\text{ref}}$, the heat flow controller compensates for the system's transient behavior by modulating $\dot{V}^c$, as depicted in Fig.~\ref{fig:control_schematic}.
The hot inlet temperature and the hot volumetric flow rate are assumed to remain nearly constant.
Heat-flow-rate regulation is achieved by designing a feedback law that directly uses a NN.
Following a transfer-learning strategy, the NN is first trained on input–output data generated with the calibrated physical model of Section~\ref{sec:physical_model}, which is used to select its architecture and hyperparameters and to provide initial network weights. 
The network is then fine-tuned on a dedicated, limited experimental data set (second-stage training). This approach transfers the knowledge embedded in the physical model to the data-driven controller, reducing the amount of experimental data required and the risk of overtraining on it \citep{5288526}.
The impact of parameter estimation uncertainty on the closed loop and of several design options is also investigated in this section, using the reference physical model and experimental results.

\subsection{Data Set for the Second-Stage Training of the Neural Network}

\begin{figure}
\centering
\includegraphics[width=\linewidth]{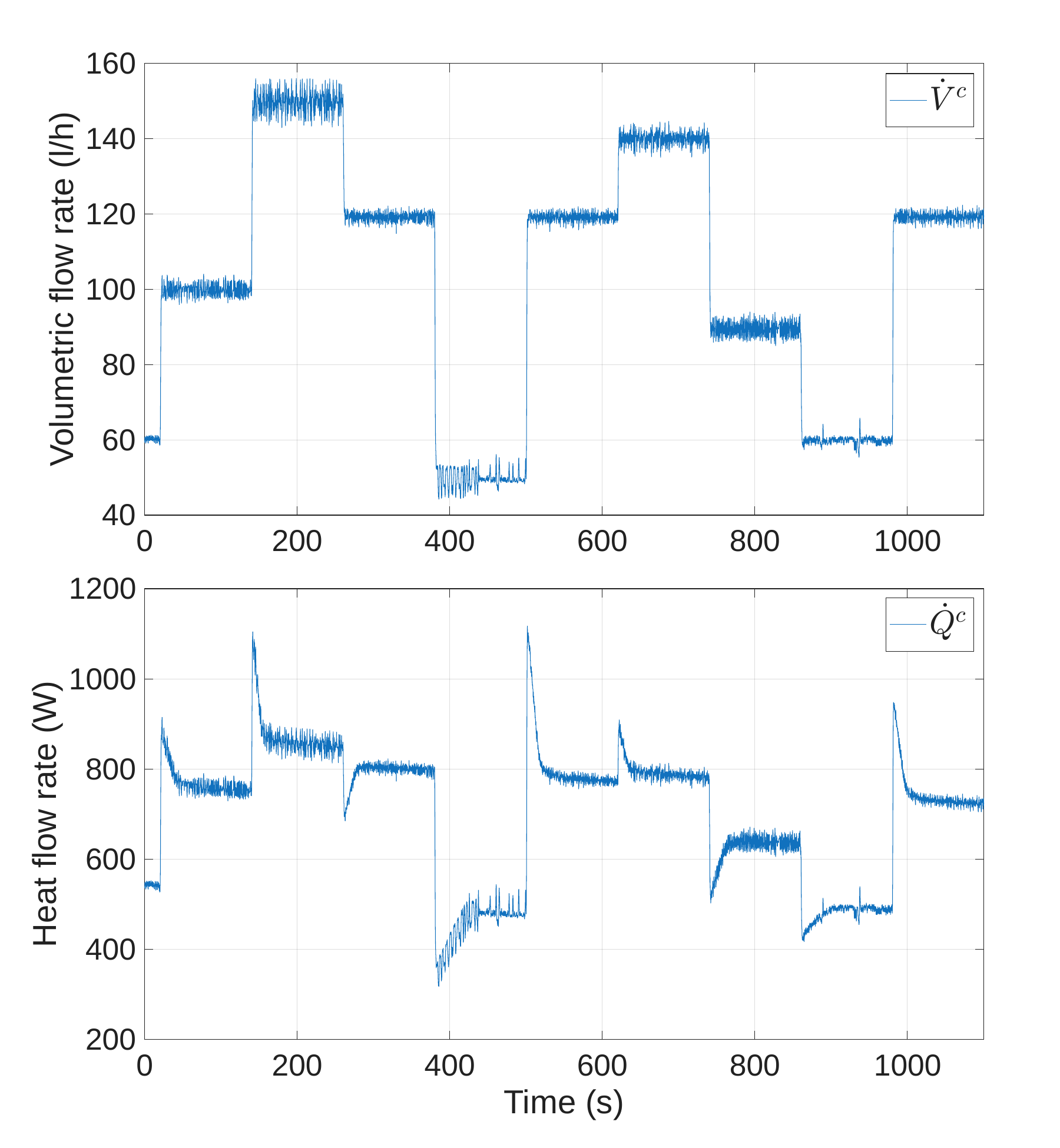}
    \caption{Training data for the NARMA-L2 Controller: input $\dot{V}^c$ (top) and output $\dot{Q}^c$ (bottom).}
    \label{fig:NARMA-L2_train}
\end{figure}

The experiment used to finalize the training of the NN on real data is similar to the one used in the previous section to evaluate the physical model (in terms of $\dot{V}^c$ fluctuations), but carried out with colder initial tank and ambient temperatures.
The reference data, shown in Fig.~\ref{fig:NARMA-L2_train}, consist of $\dot{V}^c$, regulated by the inner loop proposed in Section~\ref{sec:innerloop}, and $\dot{Q}^c$, calculated from Eq.~\eqref{eq:LMTD-Heat Rate Cold}.
The first 100\,s of data are discarded for the training, as this initial behavior does not correspond to the classical operating conditions for regulation (as discussed in the previous section).
The sequence of steps of arbitrary amplitude covers the range of expected operation, and the duration of the steps is set to capture the transient response of the heat flow rate.

\subsection{Methods: Heat Flow Regulation Using NARMA-L2 Control}

A PI control strategy is first implemented, as an off-the-shelf reference for control design. 
The baseline PI controller is chosen for its simplicity and widespread industrial application.
Its parameters are first tuned using simulations of the physical model, and then refined through experiments to suppress high-frequency oscillations.
The final controller parameters are set to 0.7 (proportional gain) and 0.05 (integral gain).

A more advanced approach is then introduced, based on feedback linearization using the nonlinear auto-regressive moving average-L2 (NARMA-L2) machine learning model \citep{Narendra97}.
The NARMA-L2 control strategy leverages NNs to model nonlinear system dynamics and is implemented in MATLAB\textsuperscript{\textregistered} Simulink\textsuperscript{\textregistered}.
Following the structure proposed by~\cite{Narendra97}, two parallel NNs model the system behavior.
Given the sampled input $\dot{V}^{c}(t)$ and output $\dot{Q}^c(t)$ at the time instant $t$, the NARMA-L2 model is expressed as:
\begin{equation}
    \dot{Q}^c(t + d t_s) = f\big(\mathbf{Y}(t), \mathbf{U}(t - t_s)\big) + g\big(\mathbf{Y}(t), \mathbf{U}(t - t_s)\big)\dot{V}^c(t)
    \label{eq:narma_l2_continuous}
\end{equation}
where $\mathbf{Y}(t) = [ \dot{Q}^c(t), \dot{Q}^c(t - t_s),  \dots, \dot{Q}^c\big(t - (n-1)t_s) ]$ is a vector containing the $n$ past sampled outputs, and $\mathbf{U}(t - t_s) = [ \dot{V}^c(t - t_s), \dot{V}^c(t - 2t_s),  \dots, \dot{V}^c\big(t - (m-1)t_s\big) ]$ contains the $m-1$ past sampled inputs.
Here, $d \geq 2$ is the relative degree of the system. The functions $f$ and $g$ are modeled using NNs.
The data-driven approximation provided by the networks captures residual nonlinearities not represented in the physical model. 
To track the reference output $\dot{Q}_{\text{ref}}(t+dt_s)$, the control input is computed by inverting Eq.~\eqref{eq:narma_l2_continuous} as:
\begin{equation}
    \dot{V}^c(t) = \frac{\dot{Q}_{\text{ref}}(t + d t_s) - f\big(\mathbf{Y}(t), \mathbf{U}(t - t_s)\big)}{g\big(\mathbf{Y}(t), \mathbf{U}(t - t_s)\big)}
    \label{eq:narma_l2_control_law}
\end{equation}

\begin{table}
    \centering \small
    \begin{tabular}{lcc}
        \toprule
        Parameter & NN1 & NN2 \\
        \midrule
        Size of the hidden layer & 5 & 20 \\
        No. of delayed plant inputs & 2 & 5 \\
        No. of delayed plant outputs & 3 & 6 \\
        \bottomrule
    \end{tabular}
    \caption{Parameters of the NARMA-L2 models for two different architectures.}
    \label{tab:NN_params}
\end{table}



Two architectures, corresponding to the levels of complexity defined by the parameters summarized in Table~\ref{tab:NN_params}, are evaluated.
The first, simpler NN architecture (NN1) comprises a single hidden layer with 5 neurons, along with 2 delayed plant inputs and 3 delayed plant outputs. The second architecture (NN2) is more complex, with 20 hidden neurons, 5 delayed inputs, and 6 delayed outputs.  These parameters are chosen according to simulations of the physical model in the domain of interest. The neural network architecture employs a hyperbolic tangent sigmoid activation function in the hidden layer and a linear activation function in the output layer. 
Both networks are then fine-tuned, starting from the weights obtained on simulated data, using the experimental data set composed of input ($\dot{V}^c$) and output ($\dot{Q}^c$) sequences (Fig.~\ref{fig:NARMA-L2_train}).  
Network training is conducted over 1,000 epochs using the Levenberg-Marquardt backpropagation algorithm.
The quality of the model fit is assessed using the normalized root mean square error (NRMSE):
\begin{equation}
\text{NRMSE fit\%} = 100 \times \left(1 - \frac{\|\mathbf{y} - \hat{\mathbf{y}}\|}{\|\mathbf{y} - \bar{\mathbf{y}}\|}\right)
\end{equation}
where $\mathbf{y}$ is the measured output data vector, $\hat{\mathbf{y}}$ is the estimated data vector, and $\|\cdot \|$ denotes the Euclidean ($L^2$) norm.

The integration of the NARMA-L2 controller into the physical model simulator increases the computation time by approximately 20$\%$. Thus, the model retains its fast computational performance, requiring only a few seconds to simulate the full experiment.

\subsection{Results and discussion}

\begin{figure}
\centering
\includegraphics[width=\linewidth]{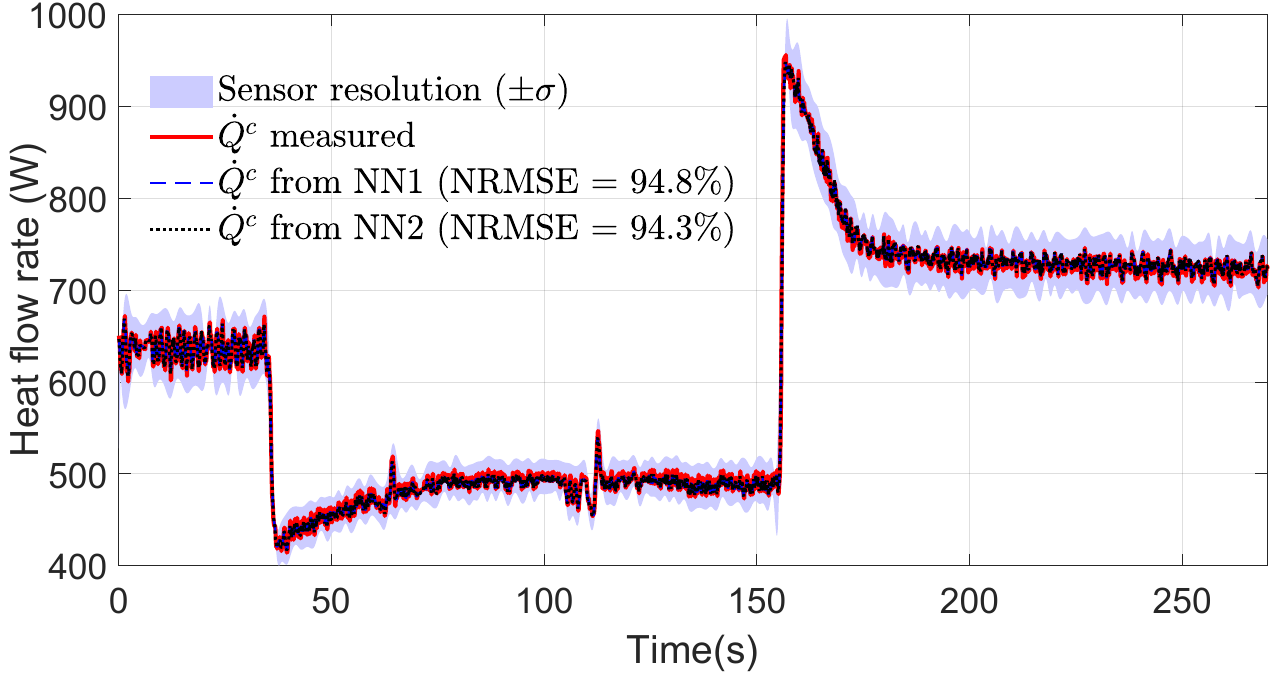}
    \caption{Comparison between the measured and NN-simulated (identified) heat flow rate. The shaded area indicates the uncertainty region.}
    \label{fig:id_figure_NRMSE}
\end{figure}

The experimental data are divided as follows: $50 \%$ for training, $25 \%$ for validation, and $25 \%$ as test data.
The identification results on the test data for NN1 and NN2 are presented in Fig.~\ref{fig:id_figure_NRMSE}. 
This figure displays the simulated output generated by the NN alongside the experimental test data, as well as the associated uncertainty region derived from measurement errors estimated based on sensor accuracy.
The simulated output is very close to the measured value and remains within the uncertainty bounds.
The resulting NRMSE values of $94.8 \%$ for NN1 and $94.3 \%$ for NN2 confirm a good fit, demonstrating that the identified model accurately captures the system's dynamics.

\begin{figure*}
\centering
\includegraphics[width=16cm]{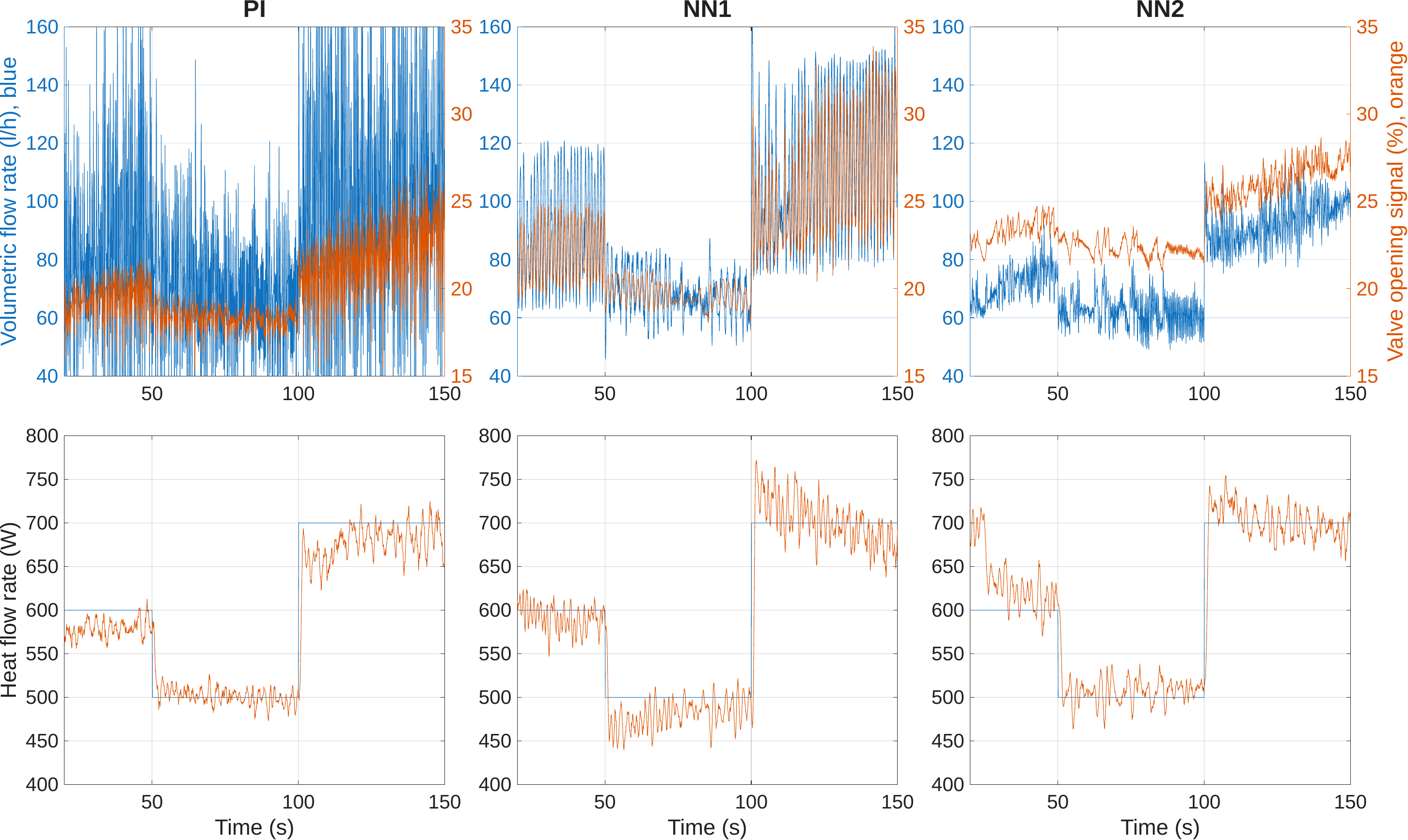}
\caption{Heat-flow-rate regulation with the proposed controllers: classical PI control (left), and reduced order (center) and more advanced (right) NARMA-L2 controls. Top: control signals produced by the controllers (blue) and resulting valve actuation (orange). Bottom: target (blue) and smoothed achieved (orange) heat flow rates.}
\label{fig:Controllers}
\end{figure*}

The tracking performance of the three controllers is presented in Fig.~\ref{fig:Controllers}. 
The top panels show the control signals $\dot{V}^c$ (blue) issued by the controllers and the corresponding actuation signals $\theta^c$ (orange, resulting from the inner loop).
The reference (blue) and achieved (orange) heat flow rates are shown in the bottom panels.
The PI controller (left column) effectively tracks the desired heat flow rate but provides a control signal with large high-frequency components (jittering), highlighting the lack of disturbance rejection and abrupt valve movements, which can result in actuator wear.
NN1 (central column) exhibits smoother control input signals than the PI controller, at the cost of a lower tracking accuracy. 
NN2 provides the best tracking performance, accompanied by significantly reduced oscillations in the control signal. 
Three metrics assess more precisely the tracking accuracy of the controllers: mean square error (MSE), standard deviation (STDV), and L2 norm of the tracking error. 
Two additional metrics evaluate control smoothness, via total variation ($\text{TV}= \frac{1}{N} \sum_{k=2}^N | U(k) - U(k-1) |$ for $N$ samples of a signal $U$), computed for both the desired volumetric flow rate $\dot{V}_{\text{des}}^{c}$ and the valve actuation signal $\theta^{c}$.
The performance of the three controllers is quantified using these metrics in Table~\ref{tab:Controllers}.
While all three controllers achieve satisfactory performance, their tracking accuracy and smoothness differ. 
Among them, NN2 yields the most accurate and stable control, with minimal tracking error and the lowest total variation.
This results in smoother control inputs and reduced actuator wear. 
By contrast, the PI controller exhibits the highest total variation in both flow rate and valve signal, indicating more aggressive and potentially fatiguing actuation, although it provides the lowest STDV.
\begin{table}
    \centering \small
    \begin{tabular}{lccc}
        \toprule
        Controller & PI & NN1 & NN2 \\
        \midrule
        MSE & 682.9 & 1332 & \textbf{609.6} \\
        STDV & \textbf{23.1} & 35.3 & 23.7 \\
        L2 norm tracking error & 906 & 1265 & \textbf{856} \\
        TV: Flow Rate (L/h) & 8.20 & 6.32 & \textbf{4.24} \\
        TV: Valve Signal (\%) & 1.45 & 0.79 & \textbf{0.28} \\
        \bottomrule
    \end{tabular}
    \caption{Performance metrics of the three controllers.}
    \label{tab:Controllers}
\end{table}

\begin{figure}
\centering
\includegraphics[width=\linewidth]{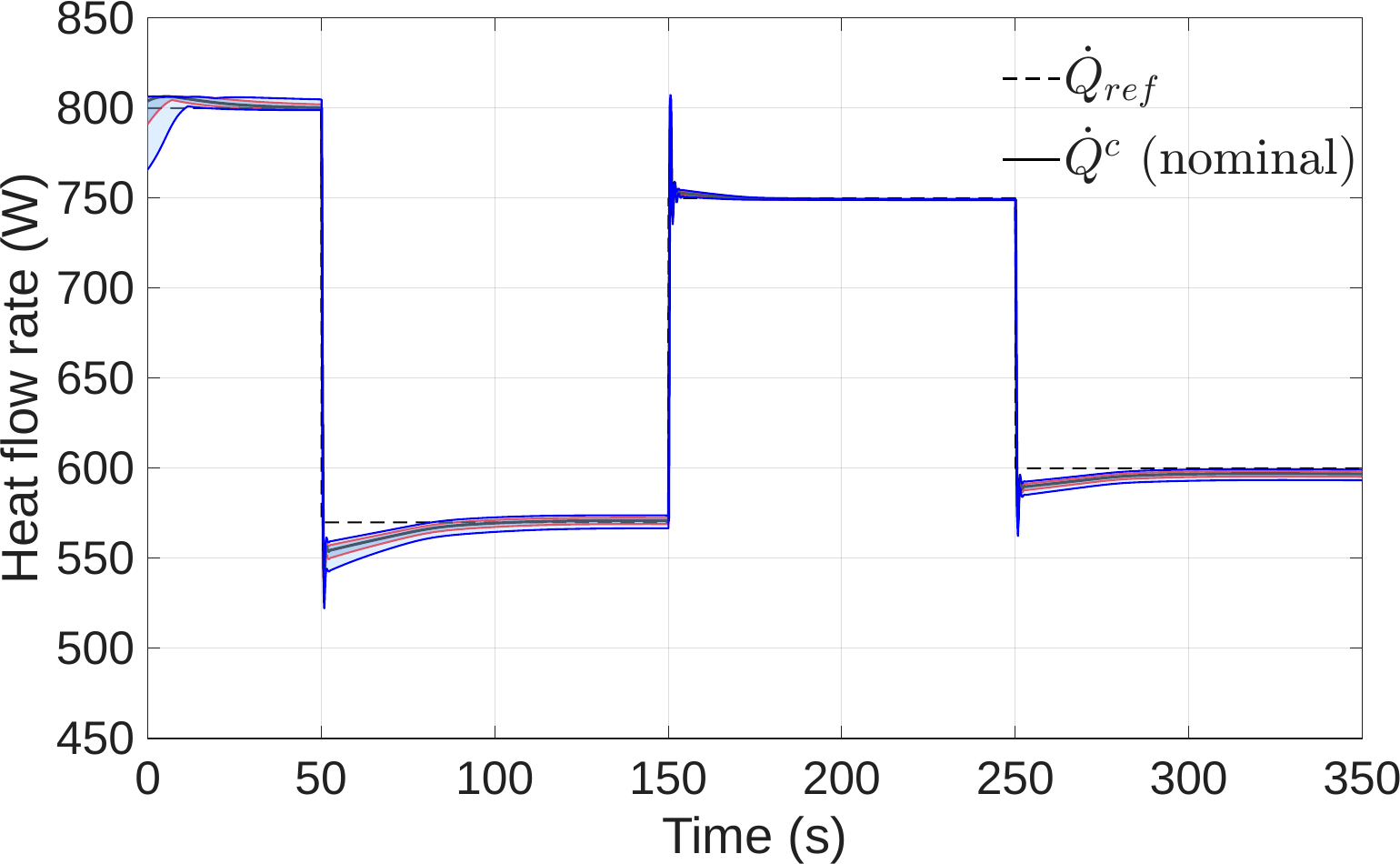}
\includegraphics[width=\linewidth]{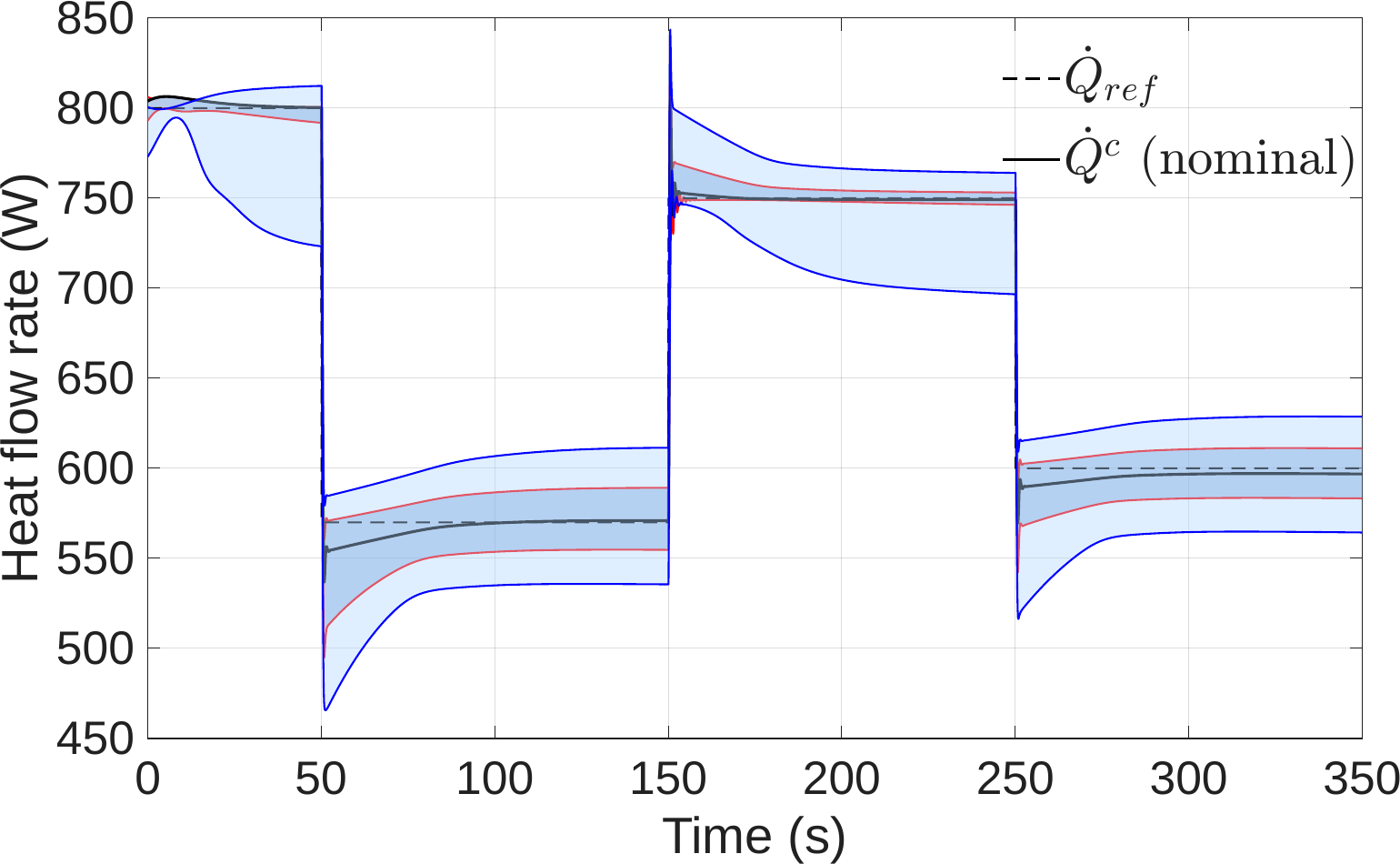}
\caption{Closed-loop performance and robustness analysis of the proposed control strategy under parameter uncertainties.
Top: simultaneous variations of $E^c$, $E^h$, and $F^h$ ($\pm 10\%$ in red and $\pm 20\%$ in blue).
Bottom: variations of the heat-transfer coefficient $h^c$ ($\pm 10\%$ in red and $\pm 20\%$ in blue).
}
\label{fig:Rob_an}
\end{figure}

Closed-loop simulations are performed to evaluate the robustness of the proposed control strategy NN2 with respect to parameter uncertainties. The results are depicted in Fig.~\ref{fig:Rob_an}. The top panel shows that, when simultaneous variations of $E^c$, $E^h$, and $F^h$ are introduced ($\pm 10\%$ and $\pm 20\%$), the closed-loop response remains closely aligned with the reference trajectory. Although small transient deviations and slight steady-state offsets appear, more pronounced at $\pm 20\%$, the controller preserves its tracking ability and achieves acceptable performance with limited overshoot and fast settling after each step change. This indicates that the control law effectively compensates for moderate inaccuracies in the system's energetic and flow-related parameters.
In contrast, the bottom panel highlights a higher sensitivity to uncertainty in $h^c$. While the $\pm 10\%$ variation still yields reasonable tracking with minor degradation, the $\pm 20\%$ case introduces more significant deviations, particularly during transients and in steady-state error levels. The spread of responses becomes noticeably larger, suggesting that $h^c$ plays the most critical role in the system dynamics and its variations are less effectively compensated by the controller. Nevertheless, the system remains stable in all cases, and the controller maintains its overall tracking capability. 

\section{Conclusion}

This work presents a novel physics-aware and data-driven method for real-time modeling and control of heat exchange dynamics in HEs.
The physical model accounts for transport phenomena, bilinearities, and nonlinear dynamics, and is calibrated using a steady-state reference data set. 
An LMTD-based methodology combined with a Dittus--Boelter correlation is employed for accurate parameterization of the heat-transfer coefficient between the two fluids.
The heat losses at the boundaries of the HE are included following a similar strategy.
This model captures transient heat-transfer mechanisms, including transport-induced delays within the HE and nonlinear coupling effects, while maintaining predictive accuracy.
It is thus particularly suitable for online regulation adaptation and diagnostics.

The control architecture integrates two nested control loops: an inner loop for precise regulation of the cold fluid's volumetric flow rate, and an outer loop for accurate tracking of the desired heat flow rate. 
The inner loop is implemented using a high-order linear ARMAX model identified from experimental data and provides a precise transient regulation of the volumetric flow rate.
The outer loop incorporates a nonlinear NN model, trained initially on the physical model and refined with experimental data.
Two control strategies for heat flow regulation are evaluated: a classical PI controller and a NARMA-L2-based NN controller. The NARMA-L2 approach significantly improves control smoothness and, with a sufficiently rich architecture, tracking performance. The best-performing architecture (NN2) exhibits lower tracking errors and reduced actuation effort compared to the PI controller, making it a promising solution for robust and efficient HE control.

Overall, the results demonstrate the effectiveness of combining physics-based modeling with machine learning techniques to achieve adaptive, real-time modeling and control of complex thermal systems.  
The proposed methodology is general and can be readily extended to other types of heat exchangers or thermal exchange processes.  
The learning algorithms involved are of moderate complexity and are suitable for real-time implementation.
The proposed approach bridges the gap between first-principles heat-transfer modeling and data-driven methods, providing both physical insight and practical predictive capability for thermal systems.

\printcredits

\section*{Declaration of competing interest}
The authors declare that they have no known competing financial interests or personal relationships that could have appeared to influence the work reported in this paper.

\section*{Data availability}
The data and most of the algorithms used to produce the figures presented in this work are available in a structured repository \citep{SkantzikasHEdata}.

\section*{Acknowledgements}
The authors acknowledge the contribution of the GIPSA-lab technical team for the design, maintenance and operation of the experimental platform.

\noindent\textit{Funding:} This work was partially supported by the CNRS Energy unit (Cellule Énergie) through the OptiDynEC project.

\appendix
\counterwithin{figure}{section}
\counterwithin{table}{section}
\counterwithin{equation}{section}

\section{Experimental Test Bench} \label{App:TestBench}


\begin{figure*}
	\centering
	\includegraphics[width=1.\linewidth]{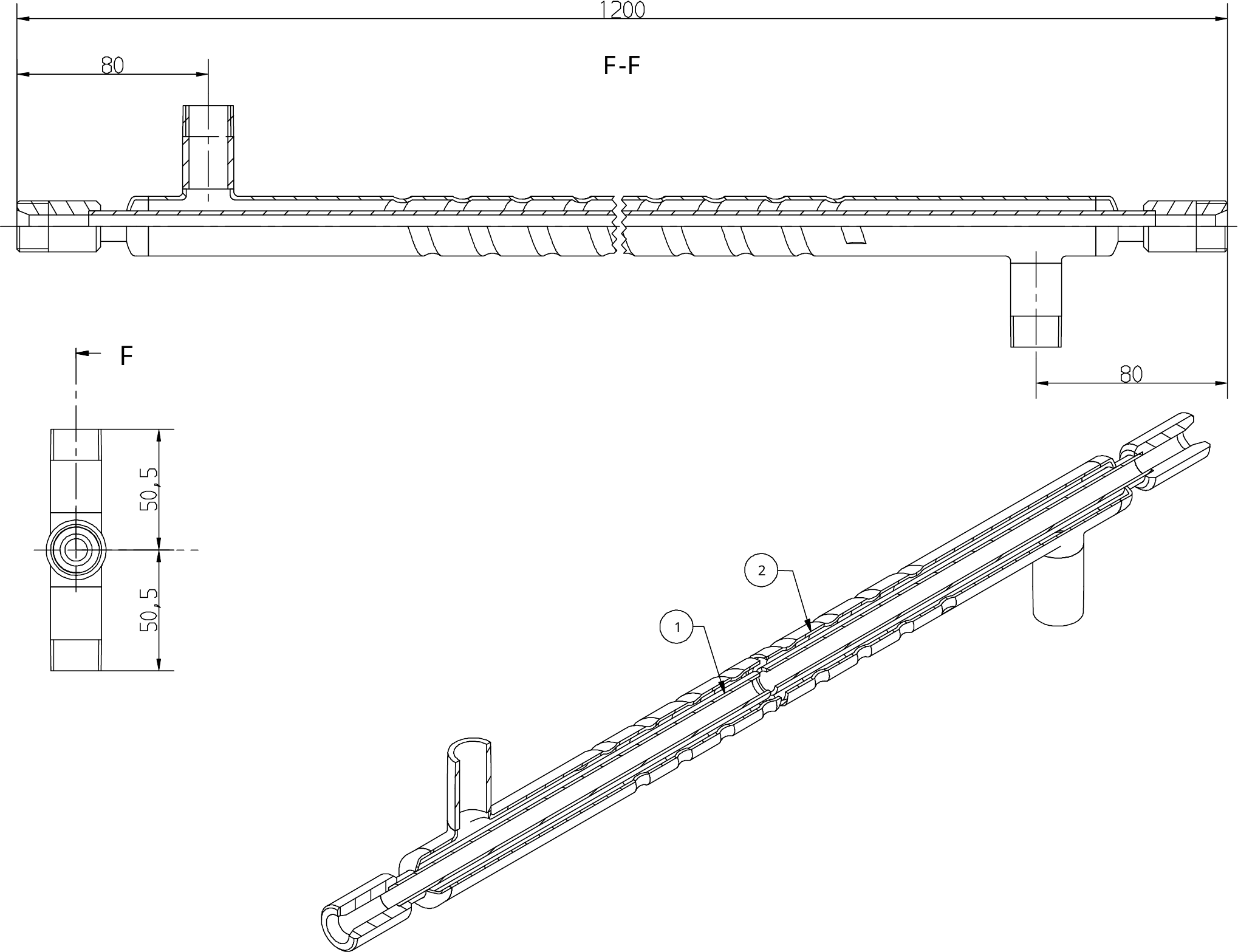}
	\caption{Perspective view and cross-sections of the Barriquand HE with the main dimensions. The inner tube (1) has dimensions $\diameter$ 12.7$\times$ 1.65 $\times$ 1140\,mm  and the outer tube (2) has dimensions  $\diameter 25\times 1.5 \times 1090$\,mm (inner diameter $\times$ thickness $\times$ length).}
	\label{fig:Gipsa_Heat_cross}
\end{figure*}

The HE system, shown in Fig.~\ref{fig:Gipsa_Heat}, consists of two large tanks (one for hot water and one for cold water), a network of pipes connecting the components, and two primary circulating pumps (\emph{Priux Home 40-25/180} high-efficiency pumps by Salmson, one for each circuit).  
Flow control is achieved via four motorized proportional valves (ASCO\textsuperscript{TM} \emph{E290}), which are controlled through a MATLAB\textsuperscript{\textregistered}-Simulink\textsuperscript{\textregistered} interface.  
The valve input is a \% of the input range (10\,V), which is converted into a voltage reference and subsequently determines the volumetric flow rate circulating through the water circuit. 
The core of the setup is a tubular double-pipe HE, depicted in Fig.~\ref{fig:Gipsa_Heat_cross}, 1.14~m long, rated at approximately 2.3~kW of thermal power, and constructed from 316L stainless steel by Barriquand Heat Exchangers\textsuperscript{TM}.  
The hot water tank is equipped with an immersed heater, which brings the fluid to a predefined temperature before circulating it through the external tube of the exchanger~\citep{Hamze_2018}.  
The cold water tank supplies fluid to the internal tube.  


The test bench includes manual valves that allow the operator to adjust flow paths, enabling operation in either parallel-flow or counter-flow mode.  
Additionally, configurations can be set to mix hot water with a colder stream or to connect the cold side directly to the water supply network.
Several sensors provide critical measurements.  
Temperature is measured at both HE boundaries (inlet and outlet for both flows) and in the tanks using \emph{PT100\textOmega \, B-Class} probes (D05979).  
Volumetric flow rates at the HE inlets are monitored using \emph{FT-110 Series} TurboFlow\textsuperscript{\textregistered} sensors.  
Pressure is measured at the HE inlet and outlet for both circuits and in the cold tank using \emph{IPS Series PS-G1000-3A4} transducers (0–1 bar range).
The system is managed via two desktop PCs: a target PC for real-time data acquisition and a host PC that communicates with it through Simulink\textsuperscript{\textregistered} xPC Target~\citep{Hamze_2018}.  
All components (sensors, valves, etc.) are connected to a National Instruments\textsuperscript{TM} \emph{NI PCI-6221} I/O board, which interfaces with the target PC.  
Communication between the PCs occurs over Ethernet.  
A real-time Simulink\textsuperscript{\textregistered} interface running on the host PC allows the user to interact with the physical system for conducting control and identification experiments.  
This setup enables full control over the valves and provides access to all acquired data, facilitating complex experimental protocols.


\section{Preliminary Experiments to Analyze the Dynamics of the Motorized Valve} \label{App:PreExp}

\begin{figure}
	\centering
	\includegraphics[width=1.\linewidth]{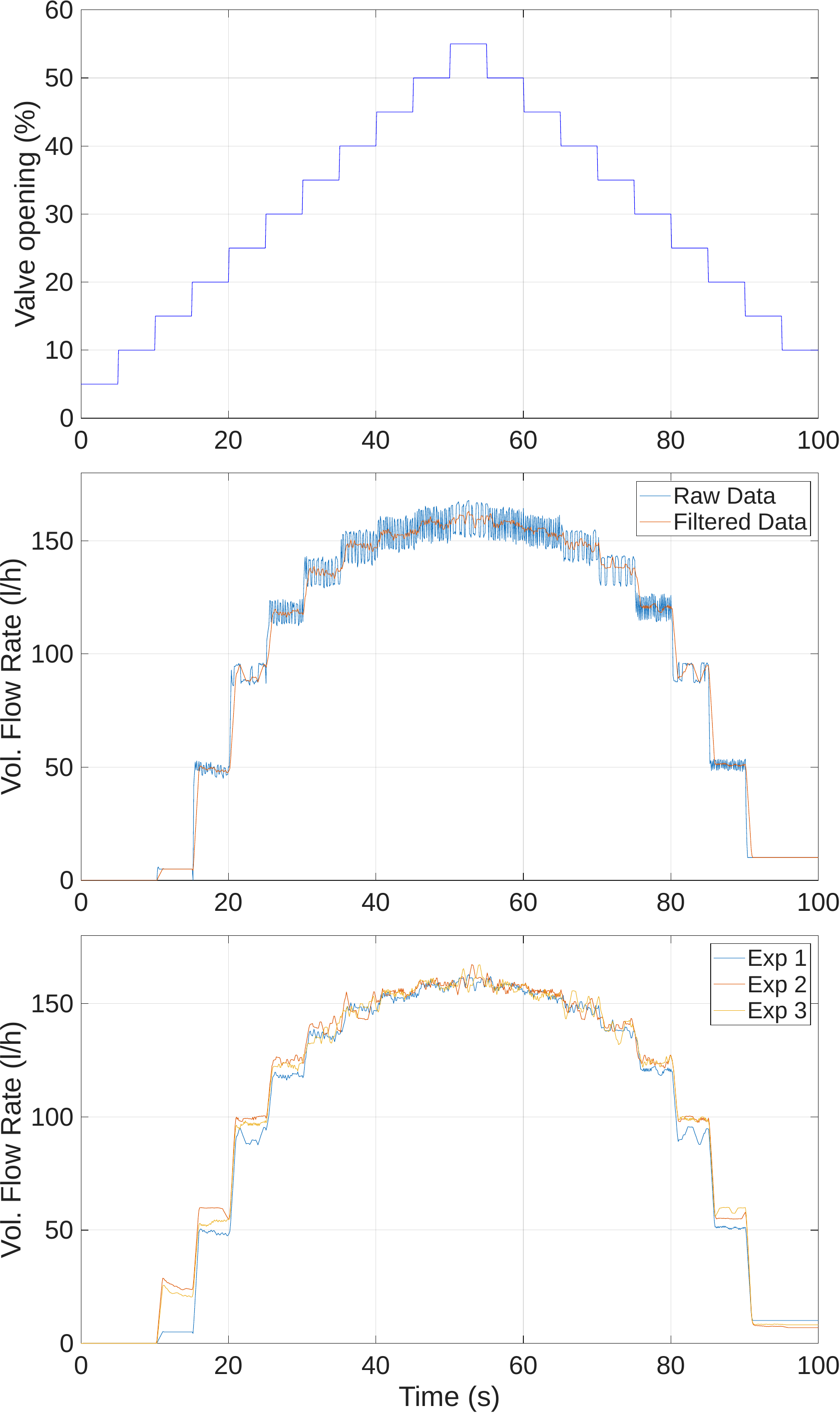}
 \caption{Preliminary experiments to evaluate the linearity of the volumetric flow rate in response to valve position modulations. Top: sequence of step inputs set for the valve opening position. Middle: raw measurements and filtered values of $\dot{V}^c$. Bottom: filtered measurements of $\dot{V}^c$ for three experiments with identical sequences of input signals.}
    \label{fig:Valve_Step}		
\end{figure}

Initial experiments conducted on the valve indicate that the maximum achievable $\dot{V}^c$ is approximately 160\,L/h. 
Considering this characteristic, the valve opening signal is set as a gradually increasing and decreasing sequence of steps, ranging from 0 to 55\,$\%$ in increments of 5\,$\%$, as depicted in Fig.~\ref{fig:Valve_Step} (top).
The raw and filtered measured values of $\dot{V}^c$ are presented in the middle, showing that high-frequency oscillations are present in the signal. The amplitude of the oscillations increases when $\dot{V}^c$ increases, as shown in Fig.~\ref{fig:Noise Amplitude}, and may reflect neglected peripheral dynamics in the flow system.
An 8$^{\textrm{th}}$-order digital moving-average filter is applied to provide a clear presentation of the signal (the control design uses raw signal only). 
The filtered signal $y_{filt}$ is obtained from the measured one $y_{meas}$ using the formula:
\begin{align} 
 y_{filt}[k]  &= \frac{1}{8}  \sum_{m=0}^{7}{y_{meas}[k-m]}  \label{EQ:IIR}
\end{align}
The variations in input/output gains for different valve positions (varying amplitude of the flow rate steps despite fixed signal steps) observed in Fig.~\ref{fig:Valve_Step} (middle) highlight the system’s pronounced nonlinear behavior, as well as the input saturation at 55\,$\%$.
Finally, three separate experiments with the same input sequence are presented in Fig.~\ref{fig:Valve_Step} (bottom), to evaluate the repeatability of the results.
While the three responses are mostly coherent, a variation of approximately 10\,L/h could be expected for an open-loop operation with a target $\dot{V}^c = 100$\,L/h, motivating the need for an efficient feedback controller.
This data set also provides information about the system transients, such as the rise time, which varies between 200 and 300\,ms, and the system delay, which is one sample (induced by the digital-to-analog converter).

\section{Extra Figures for the Heat Transport Model} \label{App:ExtraFigs}

\begin{figure}
\centering
\includegraphics[width=\linewidth]{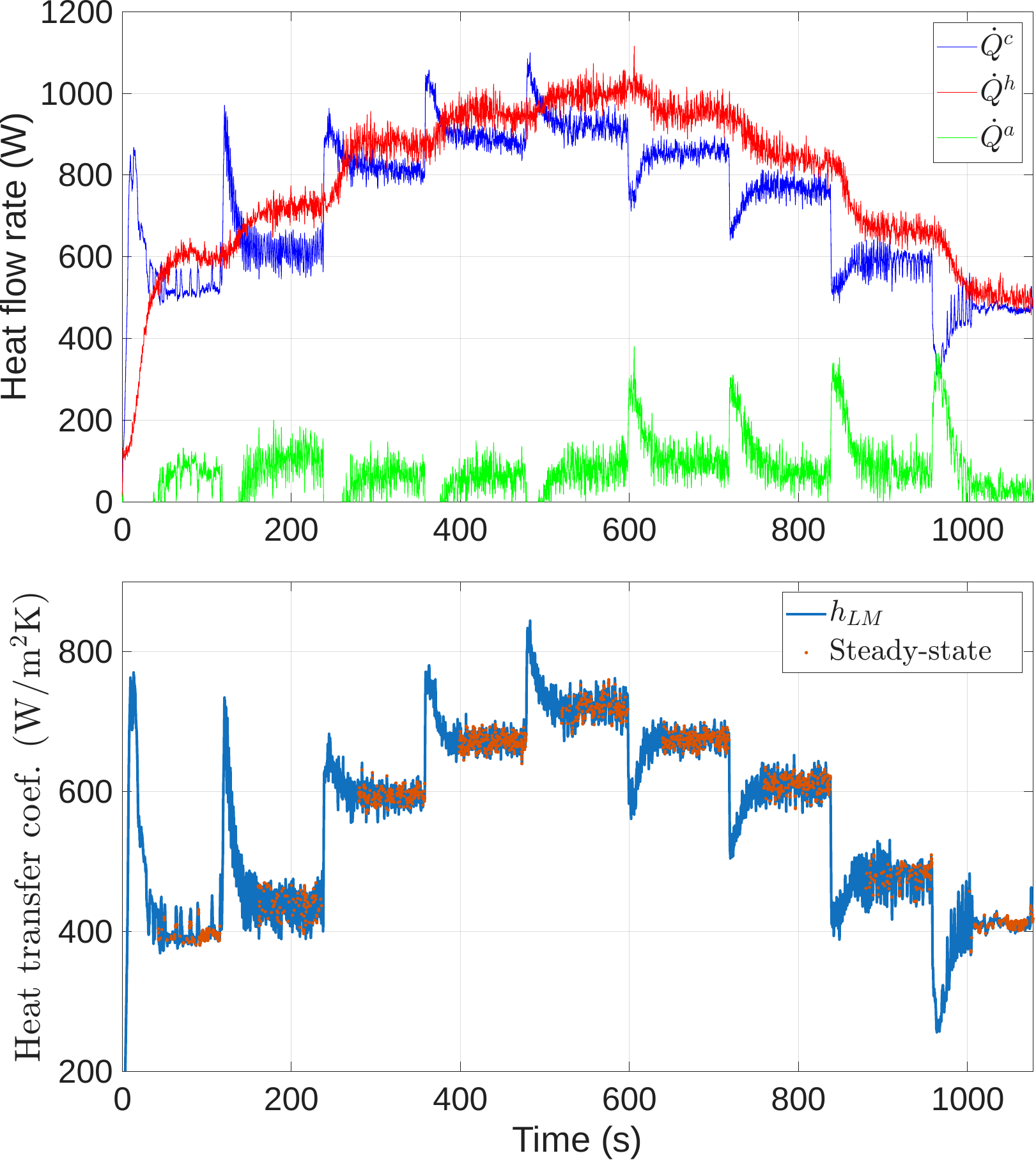}
\caption{Heat flow rates and selection of steady-state values of $h$ for the model parameterization.
Top: heat flow rates lost by the hot fluid and gained by the cold fluid and by the ambient air.
Bottom: selection of the steady-state estimates of the heat-transfer coefficient from the LMTD method.
}
 \label{fig:h estimation lmtd} 
\end{figure}

The time trends of the heat flow rates and of the heat-transfer coefficient obtained from the LMTD method are depicted in Fig.~\ref{fig:h estimation lmtd}.

\bibliography{bibHE.bib}



\end{document}